\documentclass[onecolumn,12pt,draftclsnofoot, romanappendices]{IEEEtran}
\normalsize
\pdfoutput=1

\ifCLASSINFOpdf
\else
\fi

\usepackage{acronym}
\usepackage[utf8]{inputenc} 
\usepackage[T1]{fontenc}
\usepackage{url}
\usepackage{ifthen}
\usepackage{cite}
\usepackage{graphicx}
\usepackage{hyperref}
\usepackage{amsmath}
\usepackage{amssymb}
\usepackage{bm}
\usepackage{bbm}
\usepackage{algorithmicx}
\usepackage{algorithm}
\usepackage{algpseudocode}
\usepackage{array}
\usepackage{booktabs}
\usepackage{multirow}
\usepackage{makecell}
\usepackage{mathtools}
\usepackage{mathrsfs}
\usepackage{xcolor}
\usepackage{balance}
\usepackage{tikz}
\usepackage{tikzit}
\usepackage{tikzscale}
\usepackage{subcaption}

\usetikzlibrary{arrows,positioning,shapes.geometric,decorations.pathreplacing,decorations.markings,patterns,patterns.meta,arrows.meta}

\DeclareMathOperator*{\argmax}{arg\,max}

\DeclareMathOperator{\init}{INIT}
\DeclareMathOperator{\up}{UP}
\DeclareMathOperator{\agg}{AGG}
\DeclareMathOperator{\mlp}{MLP}
\DeclareMathOperator{\relu}{ReLU}
\DeclareMathOperator{\lut}{LUT}
\newcommand{\Paren}[1]{\left(#1\right)}
\newcommand{\Br}[1]{\left[#1\right]}
\newcommand{\Brace}[1]{\left\{#1\right\}}

\newcommand{\cV}{\mathcal{V}}
\newcommand{\cG}{\mathcal{G}}
\newcommand{\cE}{\mathcal{E}}
\newcommand{\cN}{\mathcal{N}}
\newcommand{\cI}{\mathcal{I}}
\newcommand{\cF}{\mathcal{F}}
\newcommand{\bR}{\mathbb{R}}
\newcommand{\bT}[1]{\mathbb{T}_{\mathrm{#1}}}
\newcommand{\cP}{\mathcal{P}}
\newcommand{\et}{\mathrm{et}}
\newcommand{\ET}{\mathcal{ET}}
\newcommand{\bE}{\mathbb{E}}
\newcommand{\cO}{\mathcal{O}}

\acrodef{qpsk}[QPSK]{quadrature phase-shift keying}
\acrodef{bpsk}[BPSK]{binary phase-shift keying}
\acrodef{awgn}[AWGN]{additive white Gaussian noise}
\acrodef{biawgn}[BI-AWGN]{binary-input AWGN}
\acrodef{sc}[SC]{successive-cancellation}
\acrodef{scl}[SCL]{SC list}
\acrodef{bp}[BP]{belief propagation}
\acrodef{crc}[CRC]{cyclic redundancy check}
\acrodef{mdp}[MDP]{Markov decision process}
\acrodef{llr}[LLR]{log-likelihood ratio}
\acrodef{fer}[FER]{frame error rate}
\acrodef{ber}[BER]{bit error rate}
\acrodef{pm}[PM]{path metric}
\acrodef{rl}[RL]{reinforcement-learning}
\acrodef{td}[TD]{temporal difference}
\acrodef{snr}[SNR]{signal-to-noise ratio}
\acrodef{cscl}[CA-SCL]{CRC-aided SC list}
\acrodef{sarsa}[SARSA$(\lambda)$]{SARSA$(\lambda)$}
\acrodef{gnn}[GNN]{graph neural network}
\acrodef{hetgnn}[HetGNN]{heterogeneous GNN}
\acrodef{pccmp}[PCCMP]{polar-code-construction message-passing}
\acrodef{bec}[BEC]{binary erasure channel}
\acrodef{bsc}[BSC]{binary symmetric channel}
\acrodef{imp}[IMP]{iterative message-passing}
\acrodef{dql}[DQL]{deep Q-learning}
\acrodef{nn}[NN]{neural network}
\acrodef{mlp}[MLP]{multilayer perceptron}
\acrodef{genalg}[GenAlg]{genetic algorithm}
\acrodef{ai}[AI]{artificial intelligence}
\acrodef{ns}[NS]{neighborhood search}
\acrodef{drm}[dRM]{dynamic Reed-Muller}

\newtheorem{claim}{Claim}
\newtheorem{remark}{Remark}

\newcommand{\changed}[1]{\textcolor{black}{#1}}

\tikzstyle{c-froz}=[pattern=north west lines, draw=black, shape=rectangle, minimum size=1em]
\tikzstyle{c-info}=[fill=none, draw={rgb,255: red,64; green,64; blue,64}, shape=rectangle, minimum size=1em, line width=1pt]
\tikzstyle{v}=[fill=none, draw=black, shape=circle]
\tikzstyle{inputs}=[fill=black, draw=black, shape=circle, minimum size=0.5em, inner sep=0pt]
\tikzstyle{Z_empty}=[fill=none, draw=black, shape=rectangle, minimum size=21pt]
\tikzstyle{Z_fill}=[fill={rgb,255: red,191; green,191; blue,191}, fill opacity=0.5, text opacity=1, draw=black, shape=rectangle, minimum size=21pt]
\tikzstyle{Z_empty_small}=[fill=none, draw=black, shape=rectangle, minimum size=14pt]
\tikzstyle{Z_fill_small}=[fill={rgb,255: red,191; green,191; blue,191}, fill opacity=0.5, draw=black, shape=rectangle, minimum size=14pt]
\tikzstyle{emb1}=[fill={rgb,255: red,205; green,205; blue,255}, draw=black, shape=rectangle, minimum width=32pt, minimum height=1pt]
\tikzstyle{v-feat}=[fill={rgb,255: red,191; green,191; blue,191}, draw=black, shape=rectangle, xshift=-16pt]
\tikzstyle{c-feat}=[fill={rgb,255: red,128; green,128; blue,128}, draw=black, shape=rectangle, xshift=-16pt]
\tikzstyle{emb2}=[fill={rgb,255: red,105; green,105; blue,255}, draw=black, shape=rectangle, minimum width=32pt, minimum height=1pt]
\tikzstyle{emb0}=[fill={rgb,255: red,255; green,205; blue,205}, draw=black, shape=rectangle, minimum height=1pt, minimum width=32pt]
\tikzstyle{Conv}=[fill=white, draw=black, shape=rectangle, minimum width=20pt, minimum height=30pt, rounded corners=2pt]
\tikzstyle{Agg}=[fill=white, draw=black, shape=rounded rectangle, minimum width=48pt, minimum height=20pt, rounded corners=5pt]
\tikzstyle{pool}=[fill={rgb,255: red,186; green,245; blue,203}, draw=black, shape=rounded rectangle, minimum width=48pt, minimum height=20pt, rounded corners=5pt, label={center:Pool}]
\tikzstyle{cat}=[fill={rgb,255: red,227; green,245; blue,185}, draw=black, shape=rounded rectangle, minimum width=64pt, minimum height=20pt, rounded corners=5pt, label={center:Concatenate}]
\tikzstyle{input-output}=[fill=none, draw=black, shape=rounded rectangle, minimum width=48pt, minimum height=20pt, rounded corners=5pt]
\tikzstyle{mlp node}=[fill=none, draw=black, shape=circle, inner sep=1pt]
\tikzstyle{trans_state}=[single arrow, draw=black, minimum height=40pt, minimum width=1pt, inner sep=3pt, single arrow head extend=3pt]
\tikzstyle{white box}=[fill=none, draw=black, shape=rectangle, minimum height=30pt, align=center]
\tikzstyle{decision}=[fill=none, draw=black, shape=diamond, aspect=2, align=center]
\tikzstyle{tex box}=[fill=white, draw=none, shape=circle, inner sep=0]
\tikzstyle{comment}=[fill={yellow!20}, draw={black!50}, rounded corners, dashed, minimum height=20pt, shape=rectangle]
\tikzstyle{visual-emb-type}=[draw, rectangle, minimum height=20pt, minimum width=6pt, inner sep=0]

\tikzstyle{pccmp-y2v}=[->, >=stealth, dash pattern=on 3pt off 3pt, draw={rgb,255: red,128; green,128; blue,128}]
\tikzstyle{pccmp-c2c}=[->, >=stealth, dash pattern=on 1pt off 1pt]
\tikzstyle{pccmp-v2c}=[<->, >=stealth]
\tikzstyle{pccmp-v2c-new}=[->, >=stealth]
\tikzstyle{pccmp-c2v-new}=[->, >=stealth, dash pattern=on 3pt off 3pt]
\tikzstyle{plain arrow}=[->, >=stealth, line width=1pt]
\tikzstyle{plain}=[-, line width=1pt]
\tikzstyle{dashed}=[-, dash pattern=on 3pt off 3pt, draw={rgb,255: red,128; green,128; blue,128}]
\tikzstyle{solid}=[-]
\tikzstyle{vecArrow}=[-, line cap=rect, decoration={markings, mark=at position 1 with {\arrow[scale=2,thin]{open triangle 60}}}, double distance=5pt, shorten >=13pt, preaction=decorate, postaction={draw,line width=5pt, white,shorten >= 13pt}]
\tikzstyle{innerWhite}=[-, white, line width=5pt, shorten >=6pt, shorten <=1pt]
\tikzstyle{init arrow}=[->, draw={rgb,255: red,191; green,191; blue,191}, line width=1pt]
\tikzstyle{c2c}=[->, draw=blue, line width=1pt]
\tikzstyle{c2v}=[->, draw={rgb,255: red,204; green,124; blue,204}, looseness=0.5, bend left=5, line width=1pt]
\tikzstyle{v2c}=[->, draw={rgb,255: red,255; green,155; blue,255}, bend left=5, looseness=0.5, line width=1pt]
\tikzstyle{c2c-gnn}=[->, draw=blue, line width=1pt]
\tikzstyle{c2v-gnn}=[->, draw={rgb,255: red,204; green,124; blue,204}, line width=1pt]
\tikzstyle{v2c-gnn}=[->, draw={rgb,255: red,255; green,155; blue,255}, line width=1pt]
\tikzstyle{fill wrap}=[-, fill={rgb,255: red,245; green,233; blue,183}]
\tikzstyle{arrow}=[->]
\tikzstyle{bgpath}=[dashed, fill=yellow!20,rounded corners, draw=black!50]

\allowdisplaybreaks

\begin{document}

\title{Scalable Polar Code Construction for Successive Cancellation List Decoding:\\A Graph Neural Network-Based Approach}

\author{Yun~Liao,~\IEEEmembership{Graduate Student Member,~IEEE,}
        Seyyed~Ali~Hashemi,~\IEEEmembership{Member,~IEEE,}
        Hengjie~Yang,~\IEEEmembership{Member,~IEEE,}
        and~John~M.~Cioffi,~\IEEEmembership{Life~Fellow,~IEEE}
\thanks{Yun~Liao and John~M.~Cioffi are with the Department of Electrical Engineering, Stanford University, Stanford, CA 94305, USA (email: yunliao@stanford.edu; cioffi@stanford.edu).}

\thanks{Seyyed~Ali~Hashemi is with Qualcomm Technologies, Inc., Santa Clara, CA 95051, USA (email: hashemi@qti.qualcomm.com).}

\thanks{Hengjie~Yang is with Qualcomm Technologies, Inc., San Diego, CA 92121, USA (email: hengjie.yang@ucla.edu).}

}

\markboth{IEEE Transactions on Communications}%
{Submitted paper}

\maketitle

\begin{abstract}
While constructing polar codes for successive-cancellation decoding can be implemented efficiently by sorting the bit channels, finding optimal polar codes for cyclic-redundancy-check-aided successive-cancellation list~(CA-SCL) decoding in an efficient and scalable manner still awaits investigation. This paper first maps a polar code to a unique heterogeneous graph called the \emph{polar-code-construction message-passing (PCCMP)} graph. Next, a heterogeneous graph-neural-network-based \emph{iterative message-passing (IMP)} algorithm is proposed which aims to find a PCCMP graph that corresponds to the polar code with minimum frame error rate under CA-SCL decoding. This new IMP algorithm's major advantage lies in its \emph{scalability} power. That is, the model complexity is independent of the blocklength and code rate, and a trained IMP model over a short polar code can be readily applied to a long polar code's construction. Numerical experiments show that IMP-based polar-code constructions outperform classical constructions under CA-SCL decoding. In addition, when an IMP model trained on a length-$128$ polar code directly applies to the construction of polar codes with different code rates and blocklengths, simulations show that these polar-code constructions deliver comparable performance to the 5G polar codes.
\end{abstract}

\begin{IEEEkeywords}
Graph neural networks, polar code design, reinforcement learning, successive-cancellation list decoding.
\end{IEEEkeywords}

\IEEEpeerreviewmaketitle

\section{Introduction}

Polar codes, originally introduced by Ar{\i}kan in \cite{arikan}, have attracted wide interest from both academia and industry because of their capacity-achieving property for a binary-input memoryless symmetric channel under the \ac{sc} decoding. Despite being asymptotically capacity-achieving, polar codes' performance with \ac{sc} decoding is unsatisfactory for short blocklengths. The performance is improved in \cite{tal_list} by concatenating the polar code with a \ac{crc} code and adopting \ac{cscl} decoding, yet is still far from the random-coding union bound \cite{Coskun2019}. Recently, Ar\i kan improves his polar codes through the polarization-adjusted convolutional (PAC) code that closely approaches the dispersion bound of the binary-input \ac{awgn} channel under serial decoding and list decoding \cite{Arikan2019,Yao2021_entropy}. The 5G standard uses polar codes in the control channel, where short blocklength codes are required \cite{3gpp_polar}.

The polar-encoding process divides the source vector into two parts, the non-frozen bits and the frozen bits. The non-frozen bits correspond to the message, while the frozen bits are predefined values known to the decoder. The transmitter encodes source vector with the polar transformation matrix to produce a polar codeword. Polar-code construction designs the frozen set under a given channel condition and for a given decoding algorithm to minimize the \ac{fer}.

Recent research proposes multiple techniques to construct polar codes tailored for the \ac{sc} decoding. Some important examples include the use of the Bhattacharyya parameter and its variants \cite{arikan, li2013practical},  methods based on density evolution \cite{mori1,mori2}, Gaussian approximation of density evolution \cite{trifonov_GA}, channel upgrading/downgrading techniques \cite{tal_construction} for general symmetric binary-input memoryless channels, and a Monte-Carlo-based bit-channel selection algorithm that handles general channel conditions \cite{sun_MC}. Recent works \cite{Bardet2016,schurch} introduce a universal partial ordering of bit-channel reliability that leads to a polar-code construction algorithm whose complexity is sublinear in blocklength \cite{mondelli_complexity}. He \emph{et al.} propose a $\beta$-expansion construction method based on this universal partial order in \cite{beta}.

All aforementioned construction techniques rely on the premise that the information bits should be transmitted over the most reliable bit-channels to achieve the optimal error-correction performance with \ac{sc} decoding. However, experiments show that with \ac{scl} or \ac{cscl} decoding, the polar codes based upon the most reliable bit-channels' selection do not necessarily result in the best error-correction performance \cite{mondelli2014}. With \ac{scl} decoding, there may not even exist a single reliability order of bit-channels that optimizes polar-code constructions for arbitrary code rates. Nevertheless, the current 5G NR standard polar-code design uses a universal reliability order of $1024$ bit-channels \cite{3gpp.38.212}. The composition of such a universal reliability order accounts for the partial reliability order imposed by the polarization effect on bit-channels \cite{mondelli_complexity}, the distance properties \cite{mondelli2014}, and the list-decoder use. For a given channel, rigorous performance analysis for the optimal code construction under \ac{scl} decoding still remains an open problem. However, some important theoretical breakthroughs advance polar codes' understanding: \cite{Hussami2009} characterizes the polar code's minimum distance, and \cite{Bardet2016} recognizes polar codes as decreasing monomial codes. In \cite{Yao2021}, Yao \emph{et al.} developed a deterministic recursive algorithm that computes the polar codes' weight enumerating function. Recently, Co{\c s}kun and Pfister \cite{Coskun2022} analyzed the \ac{scl} decoder's required list size to approach the maximum-likelihood decoding performance. Several methods, including parity-check designs \cite{wang2016parity, zhang2018parity}, dynamic frozen bits design \cite{trifonov2015polar, trifonov2017randomized}, weight-distribution optimization \cite{Trifonov2020,Chiu2020,Rowshan2021, Miloslavskaya2021,Li2021} and altering construction patterns \cite{qin_MC, li2014rmpolar, yuan2019polar} improve polar codes' design for \ac{scl} decoding. The authors of \cite{qin_MC} propose a \ac{llr}-evolution-based construction method that swaps vulnerable non-frozen bit-channels with strong frozen bit-channels for \ac{bp} decoding of polar codes, and shows improvement with \ac{scl} decoding. \cite{li2014rmpolar} improves the polar codes' minimum distance by excluding bits corresponding to low-Hamming-weight rows in the polar transformation matrix. A recent work \cite{yuan2019polar} improves the polar codes' distance spectrum by using dynamic frozen bits that protect the low-row-weight information positions. 

Recently, \ac{ai} techniques emerge as promising tools to construct polar codes \cite{ebada2019deep, elkelesh_GA, huang_AI, Li2021_COML, liao2021construction, Huang2019}. In particular, \cite{ebada2019deep} introduces a deep-learning-based polar-code construction method for \ac{bp} decoding. Recent works \cite{elkelesh_GA} and \cite{huang_AI} propose a genetic algorithm to construct polar codes with \ac{scl} decoding. Li \emph{et al.} \cite{Li2021_COML} propose an attention-based set-to-element model to construct nested polar codes for \ac{scl} decoding. In \cite{liao2021construction}, a tabular \ac{rl} algorithm constructs polar codes for \ac{scl} decoding, and significantly reduces the training sample complexity compared to the genetic algorithm. The \ac{rl} algorithm also allows nested polar-code construction by modeling the construction process with a \ac{mdp} \cite{Huang2019}. Some of these methods can benefit from a trained model or a found solution for a slightly different target channel condition to reduce the training complexity at a new target channel. However, these methods usually require separate training for different blocklengths, code rates, and target channel conditions. In other words, these \ac{ai}-based algorithms provide satisfying polar-code constructions for the trained cases, but do not yet generalize to other code design tasks with different parameters. Moreover, their training complexity becomes prohibitively high as the blocklength increases or as the \ac{fer} decreases, making these algorithms not suitable for designing polar codes of long blocklengths.

In contrast with the aforementioned \ac{ai}-based algorithms whose complexities grow with blocklength, the \textit{\ac{gnn}} \cite{Scarselli2009,Zhou2020,Wu2021} addresses powerfully the tasks related to graph-structured data and is scalable. That is, the model complexity is independent of the graph size and the trained model readily applies to an arbitrarily large graph. A particularly useful \ac{gnn} variant is the \textit{heterogeneous GNN} \cite{zhang2019heterogeneous} that handles tasks for \textit{heterogeneous graphs}, i.e., graphs with different node types and edge types.

Inspired by \ac{gnn}'s scalability feature, this paper first maps a polar code to a unique heterogeneous graph termed as the \textit{\ac{pccmp} graph}. A heterogeneous-\ac{gnn}-based algorithm, called the \emph{\ac{imp}} algorithm, then follows. The \ac{imp} algorithm aims to find a \ac{pccmp} graph that corresponds to the polar code with minimum \ac{fer} with \ac{cscl} decoding by building the frozen set iteratively based on the target channel condition and code rate. The parameters in the \ac{imp} model are trained with the \ac{dql} method \cite{mnih2013playing}. The \ac{imp} algorithm's major advantage is its \textit{scalability}. More specifically, the number of trainable parameters in an \ac{imp} model is independent of the blocklength and code rate because all \ac{imp} operations are local on the \ac{pccmp} graph. Moreover, a trained \ac{imp} model for a short polar code directly applies to the design of a longer polar code under a different channel condition, requiring only polynomial computational complexity in blocklength. Simulations show that when the \ac{imp} model is trained and evaluated at the same blocklength and code rate, the \ac{imp} algorithm constructs polar codes that significantly outperform the Tal-Vardy constructions in \cite{tal_construction} with the \ac{cscl} decoding within the training \ac{snr} range. These \ac{imp}-based polar codes also achieve similar or lower \ac{fer} than state-of-the-art polar-code construction methods tailored for \ac{cscl} decoding. In addition, experiments verify \ac{imp}'s scalability, i.e., a single trained \ac{imp} model directly applies to various blocklengths, code rates, and different target \acp{snr}. These \ac{imp}-based polar codes achieve comparable \ac{fer} performance comparing to state-of-the-art construction methods. 

Next, Section~\ref{sec:prel} briefly introduces the preliminaries of polar codes, \acp{gnn}, and \ac{rl} systems. Then, Section~\ref{sec:pcc_process} describes the \ac{pccmp} graph design and the \ac{imp} algorithm. Section~\ref{sec:dql_details} introduces the training of the IMP model, while Section~\ref{sec:simu} presents experimental results and observations. Finally, Section~\ref{sec:conc} concludes.

\section{Preliminaries}\label{sec:prel}

\subsection{Notation}

Throughout the paper, $\log$ is in base 2; $[n]\triangleq \{0, 1, \dots, n-1\}$. A length-$n$ column vector is denoted by a bold lowercase letter, e.g., $\bm{x} \in \bR^n$, in which the $i$-th element is written as $x[i]$, $i \in [n]$; an $n\times m$ matrix is denoted by a bold uppercase letter, e.g., $\mathbf{X}\in\bR^{n\times m}$, in which the $(i,j)$-th entry of matrix $\mathbf{X}$ is given by $X[i,j]$, $i\in[n], j\in[m]$. Let $(\cdot)^\top$ represent the transpose of a vector or a matrix, $[\bm{x}_1^\top, \bm{x}_2^\top]^\top$ the vertical concatenation of $\bm{x}_1$ and $\bm{x}_2$, and $\|\bm{x}\|_2$ the $\ell_2$ norm of a vector $\bm{x}$. Sets are denoted by calligraphic letters, e.g., $\mathcal{I}$, and $|\cdot|$ represents the set's cardinality. In the description of graphs and \acp{gnn}, a directed edge from node $u$ to node $v$ is represented by $(u,v)$, and the superscript $i$ in $(\cdot)^{(i)}$ denotes the $i$-th message passing iteration. The \ac{crc} (generator) polynomial is represented in hexadecimal, in which the corresponding binary coefficients are written from the highest to the lowest order. The coefficient of the highest order bit is omitted because it is always $1$. For instance, the degree-$4$ \ac{crc} polynomial $x^4 + (x + 1)$ is written as 0x3.

\subsection{Polar Codes and \ac{sc}-Based Decoding}

$\mathcal{P}(N, K, m)$ denotes a polar code with length $N = 2^n$, rate $R = (K-m)/N$, and $m$ \ac{crc} bits, where $n \ge 0$, $0\le m \le K \le N$. A codeword $\bm{c} \in \mathbb{F}_2^N$ in $\mathcal{P}(N, K, m)$ is obtained by applying a linear transformation $\tilde{\mathbf{G}}_n$ to the source vector $\bm{u} = (u[0], u[1], \ldots, u[N - 1])^\top$ as $\bm{c} = \bm{u}^\top \tilde{\mathbf{G}}_n$, in which the polar-transformation matrix $\tilde{\mathbf{G}}_n$ is constructed from the Ar{\i}kan's polarization kernel $\mathbf{G} = \left[\begin{smallmatrix}
    1 & 0 \\
    1 & 1
\end{smallmatrix}\right]$ as 
$\tilde{\mathbf{G}}_n = \mathbf{G}^{\otimes n}$,
where $\mathbf{G}^{\otimes n}$ is the \mbox{$n$-th} Kronecker power of $\mathbf{G}$ \cite{arikan}. The source vector $\bm{u}$ contains a set $\mathcal{F}$ of $(N-K)$ frozen bits and a set $\mathcal{I}$ of $K$ non-frozen bits. The frozen bits' positions and values are known to both the encoder and the decoder, and their values are commonly set to zero. The $K$ non-frozen bits include $m$ \ac{crc} bits and $(K-m)$ information bits. If no $\ac{crc}$ is used, $m = 0$. $\mathcal{P}(N, K, m)$'s \emph{construction} selects the positions of the $(N-K)$ frozen bits and the positions of the $m$ \ac{crc} bits for a specific design channel condition. For simplicity, here, the \ac{crc} bits are appended after the $(K-m)$ information bits. Then, the construction problem simply chooses $(N-K)$ out of $N$ positions as frozen bits. Moreover, this paper considers the \ac{awgn} channel and \ac{bpsk} modulation. The \ac{snr} is defined as $\gamma \triangleq 10\log_{10}(E_s/N_0) $ dB, where $E_s$ denotes energy per transmitted symbol and $N_0$ denotes the one-sided power spectral density of the AWGN channel.

The \ac{sc} decoder and its variants use serial decoding algorithms that detect source bit $u[k]$ based on the received sequence and the previous $k$ decoded source bits. The \ac{sc} decoder assumes all previously decoded source bits are correct, which is susceptible to cascading decoding errors. The \ac{scl} decoder improves decoder performance by keeping up to $L$ most likely decoding paths in parallel \cite{tal_list}, where $L$ is the list size. When the decoding terminates, the \ac{scl} decoder selects one codeword from the candidate list either based on the likelihood, or ``pure \ac{scl} decoding'', or by \ac{crc} verification, or ``\ac{cscl} decoding''.

\subsection{Basic Concepts in Heterogeneous \acp{gnn}}

Let $\cG(\cV, \cE)$ be a graph, in which $\cV$ represents the set of nodes, and $\cE$ represents the set of directed edges, e.g., $(u,v) \in \cE$ if and only if there exists a directed edge from node $u$ to node $v$. A graph $\cG(\cV, \cE)$ is heterogeneous if the graph's nodes and edges have different types. For a heterogeneous graph $\cG(\cV, \cE)$, let $\bT{n}(u)$ and $\bT{e}((u,v))$ denote the node type for node $u\in\cV$ and the edge type for edge $(u, v)\in\cE$, respectively. For a node $u\in\cV$, its \textit{node embedding} is a vector denoted by $\bm{h}_u\in\mathbb{R}^{d}$ for some $d \ge 1$ that, after optimization, reflects the local feature and graph position of node $u$, and the structure of local graph neighborhood of node $u$ \cite[Part I]{hamilton2020graph}. Since node embeddings are often learned in an iterative manner, the notation $\bm{h}_u^{(i)}\in\mathbb{R}^{d^{(i)}}$ specifies the node embedding for node $u$ at iteration $i$, where $i\ge0$, and $d^{(i)}\ge1$ is a hyper-parameter to be specified. The initial node embedding $\bm{h}_u^{(0)}\in\mathbb{R}^{d^{(0)}}$ is typically set to a vector that captures the local features of node $u$.

A \ac{gnn} addresses tasks related to graph-structured data such as node selection, link prediction, and graph classification. The \ac{gnn}'s defining feature of a \ac{gnn} is its use of neural message passing, in which a \textit{neighborhood aggregator} exchanges each node's local messages between its neighbors, and a small \acp{nn} \textit{updates} each node's embedding \cite{gilmer2017neural}. The \ac{gnn} performs this process iteratively. 
Eventually, these final node embeddings are used to solve a graph-related task.

A heterogeneous GNN \cite{zhang2019heterogeneous} handles tasks for heterogeneous graphs. In general, for a heterogeneous graph $\cG(\cV, \cE)$, a heterogeneous GNN performs the neighborhood aggregation and update for node $u\in\cV$ according to the node type $\bT{n}(u)$ and edge type $\bT{e}((v, u))$, where $(v, u)\in\cE$. Eventually, all type-dependent updates aggregate into an updated node embedding. This work considers a simplified \ac{gnn}, in which the neighborhood aggregators and update operations only depend on the edge type $\bT{e}((v,u))$.  Formally, define the in-neighborhood for a node $u\in\cV$ with edge type $\et$ by
\begin{align}
  \cN_{\et}(u) \triangleq \Brace{v \in \cV\ |\ (v, u)\in\cE, \bT{e}((v,u)) = \et},
\end{align}
In addition, define the set of edge types associated with node $u\in\cV$ by
\begin{align}
  \ET(u)\triangleq \{\bT{e}((v,u))\ |\ (v, u)\in\cE \}.
\end{align}
The node embedding update for node $u\in\cV$ at iteration $i$ executes in three steps.
\begin{itemize}
  \item[1)] \textit{Type-wise neighborhood aggregation}: Node $u$ aggregates messages from $\cN_{\et}(u)$ for every edge type $\et\in\ET(u)$, i.e.,
    \begin{align}\label{eq:gnn_type_wise_agg_rule_hetero}
      \bm{g}_{u,\et}^{(i)} = \agg_{\et}^{(i)} \Paren{\Brace{\bm{h}_v^{(i)} \ |\ v \in \mathcal{N}_{\et}(u)}},\quad \forall \et\in\ET(u),
    \end{align}
    where $\agg_{\et}^{(i)}$ is the neighborhood aggregator for edge type $\et$ during iteration $i$.
  \item[2)] \textit{Type-wise update}: Node $u$ computes the update for every edge type $\et\in\ET(u)$ by
    \begin{align}\label{eq:gnn_update_rule_hetero}
      \bm{h}_{u,\et}^{(i)} = \up_{\et}^{(i)} \Paren{\bm{h}_u^{(i)}, \bm{g}_{u,\et}^{(i)}},\quad \forall \et\in\ET(u),
    \end{align}
    where $\up_{\et}^{(i)}$ denotes the update operation for edge type $\et$ during iteration $i$. 
  \item[3)] \textit{Local aggregation}: Node $u$ further aggregates all edge-type-dependent updates, i.e.,
    \begin{align}\label{eq:gnn_local_agg_rule_hetero}
      \bm{h}_u^{(i+1)} = \agg_{\bT{n}(u)}^{(i)} \Paren{ \Brace{ \bm{h}_{u, \et}^{(i)}\ |\ \et \in \ET(u)}},
    \end{align}
    where $\agg_{\bT{n}(u)}^{(i)}$ denotes the local aggregator for node type $\bT{n}(u)$ during iteration $i$.
\end{itemize}

\begin{figure}[t!]
    \centering
    \includegraphics{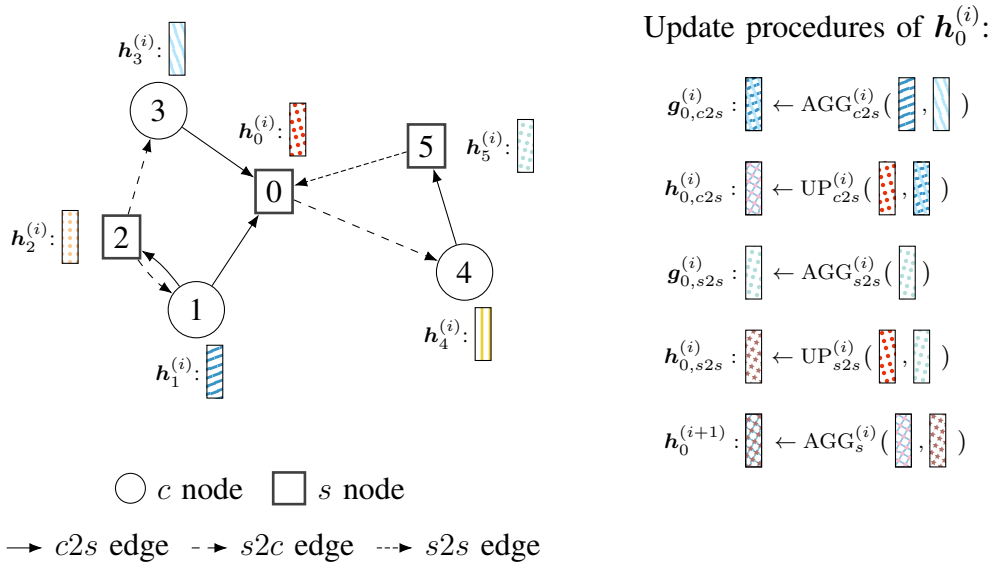}
    \caption{Node embedding update of $\bm{h}_0^{(i)}$ in a heterogeneous graph.}
    \label{fig:gnn_single_step}
    \vspace{-2em}
\end{figure}

As an example, let $\cG(\cV, \cE)$ be a heterogeneous graph with $6$ nodes and $8$ edges depicted in Fig. \ref{fig:gnn_single_step}. Each node is of one of the two node types: $\{c, s\}$, and each edge is of one of the three edge types $\{c2s, s2c, s2s\}$. Consider the update procedures of $\bm{h}_0^{(i)}$ for node $0$. According to \eqref{eq:gnn_type_wise_agg_rule_hetero} and \eqref{eq:gnn_update_rule_hetero}, the heterogeneous GNN computes $\bm{h}_{0,c2s}^{(i)}$ from embeddings at nodes $1$ and $3$ whose edges to node $0$ are of type $c2s$, and $\bm{h}_{0,s2s}^{(i)}$ from embedding at node $5$ whose edge to node $0$ is of type $s2s$. Finally, the heterogeneous GNN produces new embedding $\bm{h}_0^{(i+1)}$ by aggregating $\bm{h}_{0,c2s}^{(i)}$ and $\bm{h}_{0,s2s}^{(i)}$ using \eqref{eq:gnn_local_agg_rule_hetero}. Fig.~\ref{fig:gnn_single_step} also shows the detailed update procedure.

\begin{remark}
  In the above three steps, the aggregators in steps 1 and 3 are permutation invariant operators such as summation or averaging of incoming messages. Only the update operator in step 2 includes trainable parameters, hence requires training. As can be seen, the model complexity, defined by the number of trainable parameters, is independent of the input graph size. More importantly, the local processing feature of the heterogeneous \ac{gnn} model makes the generalization over graphs possible: a well-designed heterogeneous \ac{gnn} model trained over small heterogeneous graphs thus directly applies to significantly larger graphs with similar properties, e.g., the same set of node types and edge types, and similar local structures around the nodes.
\end{remark}

\subsection{Basics of a \ac{rl} System}

\ac{rl} is a machine learning technique where an agent learns in an interactive environment by trial and error using feedback from its own actions and experiences. \ac{rl} techniques are particularly powerful for a \ac{mdp} environment, in which the environment's state transition and feedback to the agent (reward) are conditionally independent of the agent's interaction history with the environment given the current environment state and the agent's action. A typical \ac{rl} setup for a \ac{mdp} environment is defined by the following key elements:
\begin{itemize}
  \item a set $\mathcal{S}$ representing \textit{states} of the environment;
  \item an \textit{action space} $\mathcal{A}(s)$ that specifies the set of actions that the agent can take at state $s$;
  \item a \textit{transition rule} $\Pr(s'|s,a)$, which is the probability of transitioning to state $s' \in \mathcal{S}$ at the next time step given that the agent takes action $a \in \mathcal{A}(s)$ at the current state $s$. The agent may not know the transition probability;
  \item an immediate \textit{reward} function $r(s,a,s')$ that determines the agent's received reward when the agent takes action $a$ at state $s$ and the environment transitions to state $s'$. The reward function  can be stochastic.
\end{itemize}

The interaction between the agent and the environment is an iterative process. At each time $t$, the agent senses the environment's state $s_t$ and chooses an action $a_t$ from $\mathcal{A}(s_t)$. The environment, stimulated by the agent's action, changes its state to $s_{t+1}$ and sends reward $r_{t+1}$ back to the agent. The agent accumulates the rewards as this interaction proceeds. The goal of the agent is to learn a \textit{policy} $\pi: \mathcal{S} \times \mathcal{A} \to [0,1]$ with $\pi(a|s)= \Pr(A_t = a|S_t = s)$ to maximize the expectation of the long-term \textit{return} $R$. Here, the long-term return is defined as $R \triangleq \sum_{t = 0}^T \beta^t r_{t+1}$, where $\beta \in [0,1]$ is the discount factor that describes how much the agent weights the future reward, and $T$ denotes the termination time. If a policy $\pi$ is deterministic, then, for all $s\in \mathcal{S}$, $\pi(a|s) = 1$ for some action $a$ and $\pi(a'|s) = 0$ for all $a' \neq a$. For simplicity, a deterministic policy $\pi$ is also written as $\pi(s) = a$.

\section{Polar-Code Construction based on Message Passing Graph}\label{sec:pcc_process}

For a given $N$, $K$, degree-$m$ \ac{crc} polynomial, list size $L$, and a target \ac{snr} $\gamma$, the goal is to find a $\cP(N, K, m)$ with polynomial computational complexity in $N$ and $K$, to minimize the \ac{fer} at \ac{snr} $\gamma$ under \ac{cscl} decoding with list size $L$. This work uses a \ac{gnn}-based technique to tackle this problem.

Each $\cP(N, K, \cdot)$ is first mapped to a unique heterogeneous graph, named the \ac{pccmp} graph. Then, a heterogeneous \ac{gnn}-based \ac{imp} algorithm for finding a \ac{pccmp} graph is presented. The \ac{imp} model is later trained with \ac{rl} to optimize the \ac{fer} performance of the polar codes corresponding to the found \ac{pccmp} graphs under \ac{cscl} decoding. This section focuses on the introduction of the \ac{pccmp} graph and the \ac{imp} algorithm, whereas Sec. \ref{sec:dql_details} elaborates on the training of the \ac{imp} model using \ac{rl}.


\subsection{Polar-Code-Construction Message-Passing Graph}

\begin{figure}[t!]
  \centering
  \begin{tikzpicture}[scale=0.8]
	\begin{pgfonlayer}{nodelayer}
	    \foreach \i in {0,...,3}
	         \node [style=v, label={above: {$y_{\i}$}}] (\i) at (2, {11-1.5*\i}) {};
		\node [style=c-froz, label={[xshift=-0.5em]above: $c_0$}] (4) at (5.5, 11) {};
		\node [style=c-info, label={[xshift=-0.5em]above: $c_1$}] (5) at (5.5, 9.5) {};
		\node [style=c-froz, label={[xshift=-0.5em]above: $c_2$}] (6) at (5.5, 8) {};
		\node [style=c-info, label={[xshift=-0.5em]above: $c_3$}] (7) at (5.5, 6.5) {};
		\node [style=v, label={right: variable nodes}] at (8, 10.25) {};
		\node [style=c-froz, label={right: frozen check nodes}] at (8, 9.5) {};
		\node [style=c-info, label={right: non-frozen check nodes}] at (8, 8.75) {};
		\node [style=none] (v2c-v) at (7.5, 8) {};
		\node [style=none, label={right: {$v2c$ edges}}] (v2c-c) at (8.25, 8) {};
		\node [style=none] (c2v-c) at (7.5, 7.25) {};
		\node [style=none, label={right: {$c2v$ edges}}] (c2v-v) at (8.25, 7.25) {};
		\node [style=none] (c2c-1) at (7.5, 6.5) {};
		\node [style=none, label={right: {$c2c$ edges}}] (c2c-2) at (8.25, 6.5) {};
		
	\end{pgfonlayer}
	\begin{pgfonlayer}{edgelayer}
	    \foreach \i\j in {0/4,1/4,2/4,3/4,2/6,3/6,1/5,3/5,3/7}
	        {\draw [style=pccmp-v2c-new] ([yshift=1pt] \i.east) to ([yshift=1pt] \j.west);
	        \draw [style=pccmp-c2v-new] ([yshift=-1pt] \j.west) to ([yshift=-1pt] \i.east);}
	    \foreach \i in {4,5,6}
	        {\pgfmathtruncatemacro{\jstart}{1+\i}
	        \foreach \j in {\jstart,...,7}
	            \draw [style=pccmp-c2c, bend left={10*(\j-\i)}] (\i) to (\j);}
		\draw [style=pccmp-v2c-new] (v2c-v) to (v2c-c);
		\draw [style=pccmp-c2v-new] (c2v-c) to (c2v-v);
		\draw [style=pccmp-c2c, bend left=10] (c2c-1) to (c2c-2);
	\end{pgfonlayer}
\end{tikzpicture}
  \caption{\ac{pccmp} graph for $\mathcal{P}(4,2, \cdot)$ with $\mathcal{I} = \{1,3\}$ and $\mathcal{F} = \{0,2\}$.}
  \label{fig:mp_graph_single}
  \vspace{-2em}
\end{figure}
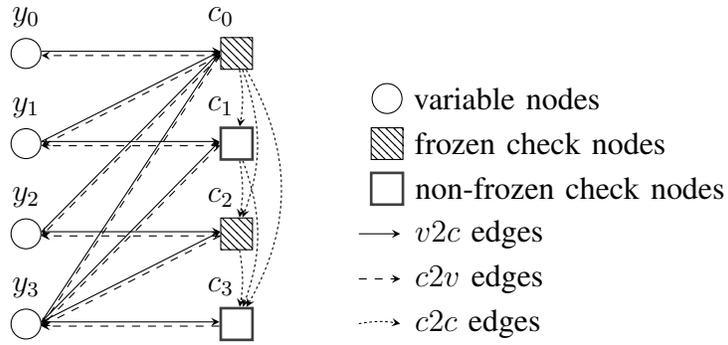

In analogy with the construction of Tanner graphs for \ac{bp} decoding, a $\mathcal{P}(N,K,\cdot)$ polar code with non-frozen set $\cI$ and frozen set $\cF$ uniquely maps to a \acf{pccmp} graph $\mathcal{G}_N(\mathcal{V}_N, \mathcal{E}_N)$. This graph is heterogeneous and is generated as follows: first, a bipartite graph is constructed with $N$ variable nodes $\mathcal{Y}_N \triangleq\{y_0, \ldots, y_{N-1}\}$ and $N$ check nodes\footnote{The term \emph{check node} is slightly abused here. Only nodes $c_i$, $i \in \mathcal{F}$ are real check nodes in \ac{bp} decoding and have predetermined values. All other nodes $c_i$ represent the non-frozen bits and need to be recovered.} $\mathcal{C}_N \triangleq \{c_0, \ldots, c_{N-1}\}$. Namely, $\mathcal{V}_N = \mathcal{Y}_N\cup \mathcal{C}_N$. For $0\le i,j\le N-1$, if $\tilde{G}_n[i,j] = 1$,  two directed edges $(y_i, c_j)$ and $(c_j, y_i)$ are added to $\mathcal{E}_N$. Second, directed edges $(c_i, c_{i'})$ are appended to $\mathcal{E}_N$ for all $0\le i< i'\le N-1$. Third, denote by $\bT{n}(v) = \mathrm{Y}$ for $v\in\mathcal{Y}_N$. Similarly, denote by $\bT{n}(c_i) = \mathrm{F}$ for $i\in\cF$ and by $\bT{n}(c_i) = \mathrm{I}$ for $i\in\cI$. Finally, the edge types are denoted by
\begin{align}
    &\bT{e}((y_j, c_i)) = v2c, \quad \text{for } i,j \in [N],~(y_j, c_i)\in\mathcal{E}_N, \\
    &\bT{e}((c_i, y_j)) = c2v, \quad \text{for } i,j \in [N],~(c_i, y_j)\in\mathcal{E}_N, \\
    &\bT{e}((c_i, c_{i'})) = c2c, \quad \text{for } i,i' \in [N],~(c_i, c_{i'})\in\mathcal{E}_N.
\end{align}
As an example, Fig.~\ref{fig:mp_graph_single} illustrates the \ac{pccmp} graph for $\mathcal{P}(4,2, \cdot)$ with $\mathcal{I} = \{1,3\}$ and $\mathcal{F} = \{0,2\}$. The rationale for the edges between check nodes clarifies after Section~\ref{subsec:imp}'s description of the \ac{imp} algorithm, and hence is stated in Remark~\ref{remark:c2c_edges}.

\begin{remark}
    As can be seen, the \ac{pccmp} graph does not rely on the CRC length $m$. The \ac{pccmp} graph's structure (i.e., nodes and connections) for $\mathcal{P}(N, K, \cdot)$, as well as the edge-types, depends only on $N$, and is independent of the code construction, i.e., the non-frozen set $\mathcal{I}$ and frozen set $\mathcal{F}$. Similar construction-independent feature can be observed in the \ac{sc}-decoding factor-graph representation \cite{arikan}, in which different constructions share the same factor graph that differ only in processing functions.
\end{remark}

\subsection{Iterative Message-Passing (IMP) Algorithm}\label{subsec:imp}
\begin{algorithm}[t!]
    \caption{\ac{imp} Algorithm}\label{alg:imp_alg}
    \hspace*{\algorithmicindent} \textbf{Input:} 
    \parbox[t]{400pt}{$N$, $K$, local feature $x_u$, $\forall u \in \cV_N$} 
    
    \hspace*{\algorithmicindent} \textbf{Output:} Selected frozen set $\cF$ and nonfrozen set $\cI$.
    \begin{algorithmic}[1]
    \State Initialize $\theta \gets 1$; \ac{pccmp} graph $\mathcal{G}_{N}$, in which $\bT{n}(u) = \mathrm{I},~\forall u \in \mathcal{C}_N$.
      \For{step $t \gets 1:N-K$}
          \State Initialize $\bm{h}_u^{(0)} \gets \init_{\bT{n}(u)}\Paren{x_u}, \forall u \in \cV_N$.
          \State $\{\bm{h}_u^{(M)} \}_{u \in \cV_N} \gets \textsc{GNN\_Processing}\Paren{\{\bm{h}_u^{(0)}\}_{u \in \cV_N}}$.
          \State $j^* \gets \textsc{Post\_Processing}\Paren{\{\bm{h}_{y_j}^{(M)}\}_{j \in [N]}, \{\bm{h}_{c_j}^{(M)}\}_{j \in [N]},\{j\}_{j \in [N] : \bT{n}(c_j) = \mathrm{I}}, \theta}$.
          \State Update $\mathcal{G}_N$ by setting $\bT{n}(c_{j^*}) \gets \mathrm{F}$.
          \State $\theta \leftarrow \theta - \frac{1}{N-K}$.
      \EndFor
      \State $\cI \gets \{j \mid j \in [N], \bT{n}(c_j)=\mathrm{I}\}$; $\mathcal{F} \gets \{j \mid j \in [N], \bT{n}(c_j)=\mathrm{F}\}$.
    \end{algorithmic}
\end{algorithm}

\begin{algorithm}[t!]
    \caption{$\textsc{GNN\_Processing}\Paren{\{\bm{h}_u^{(0)} \}_{ u \in \cV_N}}$}\label{alg:gnn_processing}
    \hspace*{\algorithmicindent} \textbf{Input:} 
    \parbox[t]{400pt}{Initial embedding $\{\bm{h}_u^{(0)} \}_{u \in \cV_N}$.} 
    
    \hspace*{\algorithmicindent} \textbf{Output:} Final embedding $\{\bm{h}_u^{(M)} \}_{u \in \cV_N}$.
    \begin{algorithmic}[1]
        \For{$i \gets 0:M-1$}
            \For{$u \in \cV_N$}
                \For{$\et \in \ET(u)$}
                    \State $\bm{g}_{u,\et}^{(i)} \leftarrow \agg_{\et} \Paren{\{\bm{h}_v^{(i)}\}_{v \in \mathcal{N}_{\et}(u)}}$; 
                    \State $\bm{h}_{u,\et}^{(i)} \leftarrow \up_{\et}^{(i)} \Paren{\bm{h}_u^{(i)}, \bm{g}_{u,\et}^{(i)}}$; 
                \EndFor
                \State $\bm{h}_u^{(i+1)} \leftarrow \agg \Paren{ \{ \bm{h}_{u, \et}^{(i)}\}_{\et \in \ET(u)}}$; 
            \EndFor
        \EndFor
    \end{algorithmic}
\end{algorithm}

\begin{algorithm}[t!]
    \caption{$\textsc{Post\_Processing}\Paren{\{\bm{h}_{y_j}\}_{j \in [N]}, \{\bm{h}_{c_j} \}_{j \in [N]}, \mathcal{L}, \theta}$}\label{alg:post_processing}
    \hspace*{\algorithmicindent} \textbf{Input:} 
    \parbox[t]{400pt}{Embeddings $\{\bm{h}_{y_j}\}$ and $\{\bm{h}_{c_j}\}$, set of candidates $\mathcal{L}\subset [N]$, auxiliary parameter $\theta$.} 
    
    \hspace*{\algorithmicindent} \textbf{Output:} Selection of $j^*$.
    \begin{algorithmic}[1]
        \State $\bar{\bm{h}}_C \gets \frac{1}{N}\sum_{j=0}^{N-1}\bm{h}_{c_j}$; $\bar{\bm{h}}_V \gets \frac{1}{N}\sum_{j=0}^{N-1}\bm{h}_{y_j}$;
        \State $\bm{g}_C \gets \tanh\Paren{\mathbf{W}_C^{\textsc{Pool}} \bar{\bm{h}}_C}$; $\bm{g}_V \gets \tanh\Paren{\mathbf{W}_V^{\textsc{Pool}} \bar{\bm{h}}_V}$;
        \State Compute $z_j \gets \mlp\Paren{\Br{\Paren{\bm{h}_{c_j}^{(M)}}^\top, \bm{g}_C^\top, \bm{g}_V^\top, \theta}^\top},~j \in \mathcal{L}$. 
        \State Select $j^* \gets \argmax_{j \in \mathcal{L}} \{z_j\}$.
    \end{algorithmic}
\end{algorithm}

The proposed polar-code construction algorithm, named the \emph{\ac{imp}} algorithm, initializes all check nodes as non-frozen, and changes one non-frozen check node to frozen in each step. The process therefore constructs $\mathcal{P}(N, K, m)$ in $N-K$ steps. In each step, the \ac{imp} algorithm iterates between a \ac{gnn}-based message-passing phase on the \ac{pccmp} graph and a post-processing phase that interprets the \ac{gnn} outputs and selects the additional frozen-check node. Algorithm~\ref{alg:imp_alg} provides the skeleton of the \ac{imp} algorithm, while Algorithms~\ref{alg:gnn_processing} and \ref{alg:post_processing} specify the detailed operations in the \ac{gnn}-processing phase, and the post-processing phase, respectively.

To exploit \ac{imp}'s full potential, this work uses small-scaled \acp{nn} in the design of the initialization operations $\init_{\bT{n}(u)}$, the update operations $\up_{\et}^{(i)}$, and the post-processing function $\textsc{Post\_Processing}(\cdot)$. The rest of this section elaborates on the design of each \ac{imp} operation.

\subsubsection{Local feature and initialization operations}

Let $x_u\in\mathbb{R}$ denote the local feature at node $u \in \cV_N$, which is set as follows: for each check node $c_j$, $j \in [N]$, $x_{c_j} = \frac{j}{N}$, which reflects the relative position of $c_j$; for variable node $y_j$, $j \in [N]$, $x_{y_j} = \gamma$, where $\gamma$ denotes the \ac{snr} defined in Section~\ref{sec:prel}. 

The initial node embedding $\bm{h}_u^{(0)}$, $u \in \cV_N$, takes the format of
\begin{equation}\label{eq:hu0}
    \bm{h}_u^{(0)} = \init_{\bT{n}(u)}(x_u) = [\bm{p}_u^\top, \bm{q}_u^\top]^\top,
\end{equation}
where $\bm{p}_u \in \bR^{d_{\mathrm{loc}}}$ is a function of $x_u$, while $\bm{q}_u \in \bR^{d_{\mathrm{type}}}$ only depends on the node type of $u$. Clearly, $d^{(0)} = d_{\mathrm{loc}} + d_{\mathrm{type}}$.

The vector $\bm{p}_u$ is computed by single-layer \acp{nn}, in which the weights and biases in the \acp{nn} are different for check nodes and variable nodes. Formally,
\begin{equation}\label{eq:pu}
    \bm{p}_u = \begin{cases}
        \tanh\Paren{\bm{w}_V^{\mathrm{loc}} x_u + \bm{b}_V^{\mathrm{loc}}}, & \text{if } \bT{n}(u) = \mathrm{Y}, \\
        \tanh\Paren{\bm{w}_C^{\mathrm{loc}} x_u + \bm{b}_C^{\mathrm{loc}}}, & \text{if } \bT{n}(u) \in \{\mathrm{I}, \mathrm{F}\},
    \end{cases}
\end{equation}
where $\bm{w}_V^{\mathrm{loc}}, \bm{b}_V^{\mathrm{loc}}, \bm{w}_C^{\mathrm{loc}}, \bm{b}_C^{\mathrm{loc}} \in \bR^{d_{\mathrm{loc}}}$ are trainable parameters. The vector $\bm{q}_u = \lut(\bT{n}(u))$, where the operation $\lut(t) \in \bR^{d_{\mathrm{type}}}$ is a lookup table that maps a categorical input $t$ to a vector with trainable elements. Note that the calculations of $\bm{h}_u^{(0)}$ at frozen check nodes and the non-frozen check nodes share the same \ac{nn} model for computing $\bm{p}_u$ by design, and they only differ in the generation of $\bm{q}_u$.

\subsubsection{\ac{gnn} processing}

In the \ac{gnn} processing phase, as detailed in Algorithm~\ref{alg:gnn_processing}, follows the rules in \eqref{eq:gnn_type_wise_agg_rule_hetero}-\eqref{eq:gnn_local_agg_rule_hetero} to perform message passing for $M$ iterations. More specifically, in each iteration, each node $u \in \cV_N$ collects and processes the incoming messages by the incoming edge types, and then combines these type-wise updates to generate the updated local embedding of $u$. In this work, the type-wise neighborhood aggregation functions $\agg_{\et}^{(i)}$ in \eqref{eq:gnn_type_wise_agg_rule_hetero} and the local aggregation functions $\agg_{\bT{n}(u)}^{(i)}$ in \eqref{eq:gnn_local_agg_rule_hetero} remain the same during all iterations $i \in [M]$. For notation simplicity, the superscript $^{(i)}$ is omitted hereinafter. Furthermore, all three types of nodes share the same local aggregation function $\agg_{\mathrm{nt}}(\cdot) = \agg(\cdot),~\forall \mathrm{nt} \in \{\mathrm{Y},\mathrm{I},\mathrm{F}\}$. The choices for $\agg_{\et}$, $\up_{\et}^{(i)}$, and $\agg$ are as follows:

\begin{itemize}
    \item \underline{Type-wise neighborhood aggregator $\agg_{\et}$}: The mean-aggregation is adopted for both edge types $c2v$ and $c2c$, i.e., 
    \begin{align}
        \agg_{\et}\Paren{\Brace{\bm{h}_v}_{v \in \cN_{\et}(u)}} = \frac{\sum_{v \in \cN_{\et}(u)} \bm{h}_v}{|\cN_{\et}(u)|},~ \et \in \{c2v, c2c\}.
    \end{align}
    The sum-aggregation is used by the edge type $v2c$ as
    \begin{equation}
        \agg_{v2c}\Paren{\Brace{\bm{h}_v}_{v \in \cN_{v2c}(u)}} = \sum_{v \in \cN_{v2c}(u)} \bm{h}_v.
    \end{equation}
    The rationale for such choices of type-wise neighborhood aggregation operators is the following: the mean-aggregation only keeps the average direction the neighboring nodes' embedding, and the result does not scale with the central node's degree. This is desired for the \ac{pccmp} graph in general because the size of in-neighborhood $\cN_{\et}(u)$ of each edge type varies from $1$ to $N$ in the graph, and maintaining the aggregated-message scale helps stabilize the training process. The aggregated message from $v2c$ edges, however, may be more helpful to the node-selection process if the degree information is included. This is because check nodes connecting to more variable nodes are more susceptible to noise, and thus might best have higher priority in the frozen-node selection process.
    \item \underline{Type-wise update operation $\up_{\et}^{(i)}$}: The update operations take the format of one convolutional layer in the GraphSAGE algorithm \cite{hamilton2017inductive}:
    \begin{equation}\label{eq:upet}
        \up_{\et}^{(i)} \Paren{\bm{h}_u, \bm{g}_{u,\et}} = 
        \mathbf{W}_{\et}^{(i)} \Br{
            \bm{h}_u^\top, \bm{g}_{u,\et}^\top}^\top + \bm{b}_{\et}^{(i)},~\et \in \ET(u),
    \end{equation}
    where $\mathbf{W}_{\et}^{(i)} \in \bR^{d^{(i+1)} \times 2d^{(i)}}$, $\bm{b}_{\et}^{(i)} \in \bR^{d^{(i+1)}}$. 
    \item \underline{Local aggregator $\agg$:} The local aggregator is given by
\begin{equation}
    \agg\Paren{\Brace{\bm{h}_{u,\et}}_{ \et \in \ET(u)}} = \relu\Paren{\frac{\sum_{\et \in \ET(u)} \bm{h}_{u,\et}}{\|\sum_{\et \in \ET(u)} \bm{h}_{u,\et}\|_2}},
\end{equation}
where $\relu(\bm{x}) \triangleq \max(\bm{x}, \bm{0})$. 
\end{itemize}

\subsubsection{Post processing}

The post processing phase computes a priority metric $z_j$ for every non-frozen check node $c_j$, $\bT{n}(c_j) = \mathrm{I}$. The non-frozen node with the largest priority metric is frozen (line 4 in Algorithm \ref{alg:post_processing}) in the updated \ac{pccmp} graph (line 6 in Algorithm~\ref{alg:imp_alg}. The post-processing operations are such that a small \ac{mlp} network is applied repeatedly at each non-frozen check node, and the size of the \ac{mlp} is independent of the blocklength $N$. The \ac{imp} algorithm with such post-processing operations thus scales to large $N$.

This phase starts with a pooling operation (lines 1 and 2 in Algorithm~\ref{alg:post_processing}) that extracts two global features $\bm{g}_C,~\bm{g}_V \in \bR^{d_{\textsc{Pool}}}$ for check nodes and variable nodes, respectively, as
\begin{align}
    &\bm{g}_C \triangleq \tanh \Br{ \mathbf{W}_C^{\textsc{Pool}} \Paren{\frac{1}{N} \sum_{j=0}^{N-1} \bm{h}_{c_j}}}, \label{eq:gc}\\
    &\bm{g}_V \triangleq \tanh \Br{ \mathbf{W}_V^{\textsc{Pool}} \Paren{\frac{1}{N} \sum_{j=0}^{N-1} \bm{h}_{y_j}}}, \label{eq:gv}
\end{align}
where $\mathbf{W}_C^{\textsc{Pool}}, \mathbf{W}_V^{\textsc{Pool}} \in \bR^{d_{\textsc{Pool}} \times d^{(M)}}$. The global features $\bm{g}_C$ and $\bm{g}_V$ are expected to provide high-level characterization about the entire graph structure and the underlying channel condition. These global features are then shared among the check nodes to assist the calculation of their final priority metrics $\{z_j\}$.

Each $z_j$ is then computed by a three-layer \ac{mlp} network with ReLU activation in all hidden layers and no (nonlinear) activation in the final (output) layer. The \ac{mlp} input concatenates the node embedding $\bm{h}_{c_j}^{(M)}$, the global features $\bm{g}_C$ and $\bm{g}_V$, and an auxiliary parameter $\theta = 1-\frac{t}{N-K} \in [0,1]$, where $t$ denotes the step index in the \ac{imp} algorithm. Formally,
\begin{equation}\label{eq:z_j}
    z_j = \mlp\Paren{\Br{\Paren{\bm{h}_{c_j}^{(M)}}^\top, \bm{g}_C^\top, \bm{g}_V^\top, \theta}^\top},~j \in [N], ~\forall \bT{n}(c_j) = \mathrm{I}.
\end{equation}
The input dimension to the \ac{mlp} is $d^{(M)} + 2 d_{\textsc{Pool}} + 1$, and the output dimension is $1$. The same \ac{mlp} applies to every non-frozen check node, and the number of trainable \ac{mlp} parameters is independent of $N$. The auxiliary parameter $\theta$ relates to the number of remaining iterations to construct the target $\mathcal{P}(N, K, m)$ polar code, and is critical in the \ac{imp} algorithm. This critical step allows the post-processing \ac{mlp} to return different priority metrics for different target code rates, even when the blocklength and the channel condition remain unchanged. These target-rate-dependent priority metrics make it possible for the \ac{imp} algorithm to produce polar-code constructions that do not depend on any global reliability ordering.

All trainable parameters in the \ac{imp} algorithm reside in (i) the initialization operations $\init_{\bT{n}(u)}$, (ii) the update operations $\up_{\et}^{(i)}$, (iii) the computation of the global features in line 2 of Algorithm~\ref{alg:post_processing} and (iv) the post-processing \ac{mlp}. As can be seen from \eqref{eq:hu0}, \eqref{eq:pu}, \eqref{eq:upet}, \eqref{eq:gc}, \eqref{eq:gv}, and \eqref{eq:z_j}, the number of trainable parameters only depends on the user-defined embedding dimensions, and is independent of $N$ and $K$. Moreover, one \ac{imp} model directly applies to arbitrary inputs $N$, $K$, and $\{x_u\}_{u \in \cV_N}$. As a result, a trained \ac{imp} model for a short polar code applies directly longer polar codes' construction.

\begin{remark}\label{remark:c2c_edges}
The edges connecting check nodes provide shortcuts on the \ac{pccmp} graph that allow each check node to receive messages directly from all of its preceding check nodes. In particular, each check node is aware of its number of frozen and non-frozen nodes that is decoded before it in \ac{cscl} decoding. Such information can affect the check nodes' priority $\Brace{z_j}$ in \eqref{eq:z_j} in the code constructions for \ac{cscl} decoding. Intuitively, in \ac{cscl} decoding, a reliable frozen bit can be helpful when there are many preceding non-frozen bits because it is likely to increase the advantage of the correct codeword's likelihood over the other candidate decoding outcomes. This influence is unique to \ac{scl} and \ac{cscl} decoding compared to the \ac{sc} decoding because of \ac{scl} and \ac{cscl}'s maintained candidate list in decoding.
\end{remark}


\subsection{Evaluation Complexity Analysis}\label{subsec:complexity}
The \ac{imp}'s construction of $\mathcal{P}(N, K, \cdot)$ requires $N-K$ steps. Each step splits into a \ac{gnn} processing phase (Algorithm~\ref{alg:gnn_processing}) and a post-processing phase (Algorithm~\ref{alg:post_processing}).

Each call of Algorithm~\ref{alg:gnn_processing} contains $M$ iterations of \ac{gnn} message passing. Each iteration consists of a type-wise local aggregation operation and a type-wise update operation. The complexity of the type-wise local aggregation depends on the number of edges in each type. The \ac{pccmp} graph contains $3^{\log N}$ $c2v$ edges, $3^{\log N}$ $v2c$ edges, and $\frac{N(N-1)}{2}$ $c2c$ edges. The complexity of the type-wise local aggregation is therefore $\cO(N^2d_{\max})$, where $d_{\max} \triangleq \max\{d^{(i)}|i \in [M+1]\}$. The type-wise update has total complexity of $\mathcal{O}(Nd_{\max}^2)$. Therefore, each call of Algorithm \ref{alg:gnn_processing} evokes $\mathcal{O}(MN^2d_{\max}+MNd_{\max}^2)$ complexity.

In Algorithm~\ref{alg:post_processing}, the computation of $\bm{g}_C$ and $\bm{g}_V$ has complexity of $\mathcal{O}(Nd_{\max} + d_{\max} d_{\mathrm{pool}})$. When $d_{\mathrm{pool}} \lesssim d_{\max}$, which is this work's setting, the complexity simplifies to $\mathcal{O}(Nd_{\max} + d_{\max}^2)$. The three-layer post processing \ac{mlp} that computes the priority metric $z_j$ for each non-frozen check node $c_j$ requires additional $\mathcal{O}(N(d_{\max}d_{\mathrm{mlp}}+d_{\mathrm{mlp}}^2))$ operations, where $d_{\mathrm{mlp}}$ is an upper bound of the \ac{mlp}'s hidden layer widths. A reasonable choice of the hidden layer sizes of the post-processing \ac{mlp} satisfies $d_{\mathrm{mlp}} = \mathcal{O}(d_{\max})$. In this case, the \ac{mlp} computational complexity simplifies to $\mathcal{O}(Nd_{\max}^2)$, and the total complexity of Algorithm~\ref{alg:post_processing} is $\mathcal{O}(Nd_{\max}^2)$.

Therefore, \ac{imp}'s total complexity for constructing $\mathcal{P}(N, K, \cdot)$ is $\mathcal{O}((N-K) (M N(Nd_{\max}+d_{\max}^2) + Nd_{\max}^2)) = \mathcal{O}(M(N-K)(N^2d_{\max} + Nd_{\max}^2))$. In practice, however, $M$ is typically a small constant no greater than $10$.

\begin{remark}
    The \acs{imp} algorithm can be potentially simplified from many aspects. One way is to use sparse $c2c$ edges (e.g., only connect adjacent check nodes) in the \acs{pccmp} graph such that the number of $c2c$ edges does not dominate the complexity analysis. Besides, the aggregation operation for all $c2v$ and $v2c$ edges can be simplified using the polar-encoding structure and achieve $\cO(N \log N d_{\max})$ complexity. These two modifications reduce the complexity of Algorithm~\ref{alg:gnn_processing} to $\cO(MN d_{\max}(\log N+ d_{\max}))$.
    
    Another way is to leverage the known bit-channel reliability from classical polar-code construction methods to freeze some check nodes in the initial stage of the \acs{imp} algorithm, such that the number of \acs{imp} steps is close to $\cO(1)$ instead of $N-K$. Applying all aforementioned modifications, the overall evaluation complexity can be potentially reduced to $\cO(Md_{\max}N(\log N+ d_{\max}))$. 
\end{remark}

\section{\ac{imp} Model Training}\label{sec:dql_details}

The training of an \ac{imp} model optimizes the trainable parameters specified in Section~\ref{sec:pcc_process} such that the \ac{imp} algorithm finds polar-code constructions with low \ac{fer} under \ac{scl} decoding.
To train an \ac{imp} model, observe that given the \ac{pccmp} graph at the end of step $t-1$ in Algorithm \ref{alg:imp_alg}, the selection of $j^*$ at step $t$ and consequently the update of the \ac{pccmp} graph are independent of the \ac{pccmp} graphs in the previous $t-2$ steps. Motivated by this crucial observation, this work models the \ac{imp} algorithm as a \ac{mdp}, with which the \ac{imp} model can be effectively learned by \ac{rl} tools.

\subsection{\ac{imp}-Based Polar-Code Construction as a \ac{mdp}}\label{subsec:mdp}

As suggested by Algorithm~\ref{alg:imp_alg}, the \ac{imp} algorithm's construction of $\mathcal{P}(N, K, m)$ requires $N-K$ iterations of frozen-bit selection. The \ac{imp}'s evolution maps to the following \ac{mdp} environment defined on the \ac{pccmp} graphs:
\begin{itemize}
    \item \textit{State}: a state $s_t$ is defined as the \ac{pccmp} graph at time $t$, i.e., $s_t = \mathcal{G}_{N,t}$, in which the set of nodes is denoted by $\mathcal{V}_{N,t} = \mathcal{Y}_{N} \cup \mathcal{C}_{N,t}$. Define $\mathcal{I}_{N,t} \triangleq \Brace{j | \bT{n}\Paren{c_j} = \mathrm{I}, c_j \in \mathcal{C}_{N,t}}$ as the set of information bits corresponding to the \ac{pccmp} graph $\mathcal{G}_{N,t}$.
    \item \textit{Action}: an action $a_t$ is the index of the selected check node that will be set to frozen at time step $t$. The set of available actions at $s_t = \mathcal{G}_{N,t}$ is $\mathcal{A}(s_t) = \mathcal{I}_{N,t}$.
    \item \textit{Rules}: the agent interacts with the environment in an episodic manner. To construct a $\mathcal{P}(N, K, m)$ polar code, each episode starts with a \ac{pccmp} graph $s_0$ with $\mathcal{I}_{N,0} = [N]$. Within an episode, if the agent takes action $a_t$ at state $s_t$, then the state transitions to $s_{t+1}$ from $s_t$ by setting $\bT{n}(c_{a_t}) = \mathrm{F}$. Note that $\mathcal{I}_{N,t+1} = \mathcal{I}_{N,t} \setminus\{a_t\}$.
    \item \textit{Termination}: an episode terminates in $N-K$ time steps.
    \item \textit{Reward}: the reward $r_{t+1}$ measures the reduction in the logarithm of \ac{fer} after setting the bit $a_t$ to frozen under \ac{cscl} decoding. Specifically, 
    \begin{equation}\label{eq:instant_reward}
        r_{t+1} = \log \Br{P_{e,t}\Paren{\mathcal{I}_{N,t}, N, m, L, \gamma}} - \log \Br{P_{e,t}\Paren{\mathcal{I}_{N,t+1}, N, m, L, \gamma}},
    \end{equation} 
    where $P_{e,t}\Paren{\mathcal{I}_{N,t'}, N, m, L, \gamma}$ is the Monte-Carlo simulated \ac{fer} at time $t$ of $\mathcal{P}\big(N, |\mathcal{I}_{N,t'}|, m \big)$ with non-frozen set $\mathcal{I}_{N,t'}$ under the \ac{cscl} decoding with list size $L$ at channel \ac{snr} $\gamma$.
\end{itemize}

The training goal is to learn a deterministic policy $\pi$ that selects an appropriate action $a_t$ at each state $s_t$ to maximize each episode's expected cumulative return.

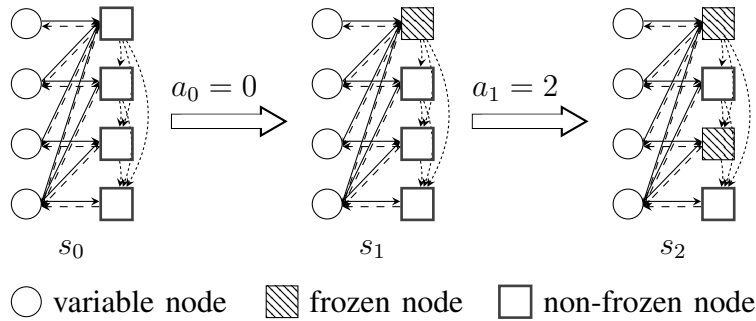
\begin{figure}
    \centering
    \begin{tikzpicture}
	\begin{pgfonlayer}{nodelayer}
	    \foreach \i in {0,...,3}
	        {
	         \pgfmathtruncatemacro{\vlabel}{4+\i}
	         \pgfmathtruncatemacro{\clabel}{8+\i}
             \node [style=v]  (\vlabel) at (0,2.5-0.8*\i) {};
             \node [style=c-info]  (\clabel) at (1.2,2.5-0.8*\i) {};}
             
        \node [style=none] at (0.6, -0.5) {$s_0$}; 
        
        \node [style=none] (t1) at (2, 1.2) {};
        \node [style=none] (t2) at (3.5, 1.2) {};
        \node [style=none, label={above:{$a_0=0$}}] at (2.5, 1.3) {};
        
        \foreach \i in {0,...,3}
	        {
	         \pgfmathtruncatemacro{\vlabel}{16+\i}
             \node [style=v]  (\vlabel) at (4,2.5-0.8*\i) {};}
        \node [style=c-froz]  (20) at (5.2,2.5) {};
        \foreach \i in {1,2,3}
            {\pgfmathtruncatemacro{\clabel}{20+\i}
             \node [style=c-info]  (\clabel) at (5.2,2.5-0.8*\i) {};}
             
        \node [style=none] at (4.6, -0.5) {$s_1$}; 
        
        \node [style=none] (t3) at (6, 1.2) {};
        \node [style=none] (t4) at (7.5, 1.2) {};
        \node [style=none, label={above:{$a_1=2$}}] at (6.5, 1.3) {};
        
        \foreach \i in {0,...,3}
	        {
	         \pgfmathtruncatemacro{\vlabel}{28+\i}
             \node [style=v]  (\vlabel) at (8,2.5-0.8*\i) {};}
        \node [style=c-froz]  (32) at (9.2,2.5) {};
        \node [style=c-info]  (33) at (9.2,1.7) {};
        \node [style=c-froz]  (34) at (9.2,0.9) {};
        \node [style=c-info]  (35) at (9.2,0.1) {};
        
        \node [style=none] at (8.6, -0.5) {$s_2$}; 
        
		\node [style=v, label={right: variable node}] at (0, -1.2) {};
		\node [style=c-froz, label={right: frozen node}] at (3.4, -1.2) {};
		\node [style=c-info, label={right: non-frozen node}] at (6.5, -1.2) {};
        
	\end{pgfonlayer}
	
	\begin{pgfonlayer}{edgelayer}
	    \foreach \i\j in {0/0, 1/0, 2/0, 3/0, 1/1, 3/1, 2/2, 3/2, 3/3} 
	        {\pgfmathtruncatemacro{\vlabel}{4+\i}
	         \pgfmathtruncatemacro{\clabel}{8+\j}
	         \draw [style=pccmp-v2c-new] ([yshift=1pt] \vlabel.east) to ([yshift=1pt]\clabel.west);
	         \draw [style=pccmp-c2v-new] ([yshift=-1pt] \clabel.west) to ([yshift=-1pt] \vlabel.east);}
	    \foreach \i\j in {0/1, 0/2, 0/3, 1/2, 1/3, 2/3}
	        {\pgfmathtruncatemacro{\startlabel}{8+\i}
	         \pgfmathtruncatemacro{\endlabel}{8+\j}
	         \draw [style=pccmp-c2c, bend left={10*(\j-\i)}] (\startlabel) to (\endlabel);}
	    \draw [style=vecArrow] (t1) to (t2);
		\draw [style=innerWhite] (t1) to (t2);
		
	    \foreach \i\j in {0/0, 1/0, 2/0, 3/0, 1/1, 3/1, 2/2, 3/2, 3/3} 
	        {\pgfmathtruncatemacro{\vlabel}{16+\i}
	         \pgfmathtruncatemacro{\clabel}{20+\j}
	         \draw [style=pccmp-v2c-new] ([yshift=1pt] \vlabel.east) to ([yshift=1pt]\clabel.west);
	         \draw [style=pccmp-c2v-new] ([yshift=-1pt] \clabel.west) to ([yshift=-1pt] \vlabel.east);}
	    \foreach \i\j in {0/1, 0/2, 0/3, 1/2, 1/3, 2/3}
	        {\pgfmathtruncatemacro{\startlabel}{20+\i}
	         \pgfmathtruncatemacro{\endlabel}{20+\j}
	         \draw [style=pccmp-c2c, bend left={10*(\j-\i)}] (\startlabel) to (\endlabel);}
	    \draw [style=vecArrow] (t3) to (t4);
		\draw [style=innerWhite] (t3) to (t4);
	
	    \foreach \i\j in {0/0, 1/0, 2/0, 3/0, 1/1, 3/1, 2/2, 3/2, 3/3} 
	        {\pgfmathtruncatemacro{\vlabel}{28+\i}
	         \pgfmathtruncatemacro{\clabel}{32+\j}
	         \draw [style=pccmp-v2c-new] ([yshift=1pt] \vlabel.east) to ([yshift=1pt]\clabel.west);
	         \draw [style=pccmp-c2v-new] ([yshift=-1pt] \clabel.west) to ([yshift=-1pt] \vlabel.east);}
	    \foreach \i\j in {0/1, 0/2, 0/3, 1/2, 1/3, 2/3}
	        {\pgfmathtruncatemacro{\startlabel}{32+\i}
	         \pgfmathtruncatemacro{\endlabel}{32+\j}
	         \draw [style=pccmp-c2c, bend left={10*(\j-\i)}] (\startlabel) to (\endlabel);}
	\end{pgfonlayer}
\end{tikzpicture}
    \caption{\ac{rl} setup of the \ac{imp}-based polar-code construction for $\mathcal{P}(4,2, \cdot)$.}
    \label{fig:pcc_mdp}
    \vspace{-2em}
\end{figure}

At the end of each episode, $\mathcal{I}_{N,N-K}$ has exactly $K$ distinct elements in $[N]$, corresponding to a valid construction for $\mathcal{P}(N, K, m)$. Fig.~\ref{fig:pcc_mdp} shows an episode of the agent's interaction with the environment for constructing $\mathcal{P}(4,2, \cdot)$.

With \eqref{eq:instant_reward}, each episode's cumulative return with design \ac{snr} $\gamma$ is
\begin{align}
    R = \sum_{t=0}^{N-K-1} \beta^t r_{t+1} = \sum_{t=0}^{N-K-1} \beta^t\big\{\log\Br{P_{e,t}\Paren{\mathcal{I}_{N,t}, N, m, L, \gamma}} - \log\Br{P_{e,t}\Paren{\mathcal{I}_{N,t+1}, N, m, L, \gamma}}\big\}. \notag
\end{align}
When $\beta = 1$, the expected cumulative return becomes
\begin{align}
    \mathbb{E}\Br{R} &= \sum_{t=0}^{N-K-1}  \bE\big[\log \Br{P_{e,t}\Paren{\mathcal{I}_{N,t}, N, m, L, \gamma}}\big] - \bE\big[\log \Br{P_{e,t}\Paren{\mathcal{I}_{N,t+1}, N, m, L, \gamma}}\big]  \\
    &= \bE\big[\log \Br{P_{e,0}\Paren{\mathcal{I}_{N,0}, N, m, L, \gamma}}\big] - \bE\big[\log \Br{P_{e,N-K-1}\Paren{\mathcal{I}_{N,N-K}, N, m, L, \gamma}}\big], \label{eq:return}
\end{align}
where \eqref{eq:return} follows from $\bE\big[\log [P_{e, t_1}(\mathcal{I}, N, m, L, \gamma)]\big] = \bE\big[\log [P_{e, t_2}(\mathcal{I}, N, m, L, \gamma)]\big]$ for any $t_1, t_2 \in [N-K]$. The first term in \eqref{eq:return} is determined by the channel's condition and the decoding algorithm; it is independent of the agent's policy $\pi$. Therefore, the policy that maximizes the expected cumulative return automatically minimizes the second term in \eqref{eq:return}. Equivalently, the policy finds a set of non-frozen bits $\mathcal{I}_{N,N-K}$ that minimizes the \ac{fer} under the \ac{cscl} decoding with list size $L$ for the target $\mathcal{P}(N, K, m)$ with design \ac{snr} $\gamma$.

With the \ac{mdp} environment, the \ac{imp} model's training uses \ac{dql} \cite{mnih2013playing} to find a return-maximizing policy.

\subsection{Training Strategies}\label{subsec:train_strategy}

Let $\Theta$ represent the set of all trainable parameters in the IMP model. Since Section~\ref{subsec:imp}'s \ac{imp} model design and the parameter sizes are independent of the \ac{pccmp} graph size and target code's specifications $(N, K,m,\gamma,L)$, the parameters $\Theta$ can be trained so that a single \ac{imp} model provides good polar codes in various code-design scenarios. With the same $\Theta$, the \ac{imp} model generates different polar codes for different input parameters $N$, $K$, and $\gamma$. To learn such parameters $\Theta$, the training can be explicitly performed over a wide range of input parameters $(N, K, \gamma)$. This paper, however, only focuses on the strategy in which the training is conducted across various $\gamma$'s with design parameters $(N, K, m, L)$ being constant. Specifically, the design \ac{snr} is sampled uniformly at random from range $[\gamma_{\min}, \gamma_{\max}]$ in each training episode. A good selection of range $[\gamma_{\min}, \gamma_{\max}]$ covers or largely overlaps the \ac{snr} range of interest and the \ac{fer} roughly ranges between $10^{-1}$ and $10^{-5}$.

Let $\Theta^*$ represent the parameters of the \ac{imp} model after training over $[\gamma_{\min}, \gamma_{\max}]$. Denote by $\mathcal{I}(N, K, m, L, \gamma; \Theta^*)$ the non-frozen set identified by this \ac{imp} model for $\mathcal{P}(N, K, m)$ at design \ac{snr} $\gamma$ under \ac{cscl} decoding with list size $L$. During evaluation, the \ac{imp} algorithm provides $\mathcal{I}(N, K, m, L, \gamma_{\mathrm{eval}}; \Theta^*)$ for each evaluation \ac{snr} point $\gamma_{\mathrm{eval}}$. The \ac{imp} model trained by this strategy is referred to as ``\ac{imp}-$[\gamma_{\min}, \gamma_{\max}]$-L$L$'' in Section~\ref{sec:simu}. When an \ac{imp} model trained on a particular pair of $(N,K)$ is directly applied to a different blocklength or code rate, the trained \ac{imp} model is labeled as ``\ac{imp}-$[\gamma_{\min}, \gamma_{\max}]$-L$L$-N$N$-K$K$'' in Section~\ref{sec:simu} for better clarity.

To further improve the performance of the trained model for each evaluation \ac{snr}, the learned parameters $\Theta^*$ are further fine-tuned by feeding a small number of additional training episodes with $\gamma = \gamma_{\mathrm{eval}}$. During this evaluation, the construction for each $\gamma_{\mathrm{eval}}$ is given by $\mathcal{I}(N, K, m, L, \gamma_{\mathrm{eval}}; \Theta^*_{\gamma_{\mathrm{eval}}})$, where $\Theta^*_{\gamma_{\mathrm{eval}}}$ represents the parameter values after fine-tuning at design \ac{snr} $\gamma_{\mathrm{eval}}$. These fine-tuned models are referred to as ``\ac{imp}-fine-tuned'' in Section~\ref{sec:simu}.

\subsection{Training Complexity Analysis}
The complexity of training an \ac{imp} model on $\mathcal{P}(N, K, m)$ using \ac{cscl} decoding with list size $L$ consists of three parts: (i) the complexity of running the \ac{imp} algorithm to decide actions; (ii) the complexity of generating the reward in each step; and (iii) the complexity of optimizing the parameters of the \ac{imp} model. The overall training complexity is linearly dependent on the number of training episodes, denoted by $T_{\mathrm{train}}$.

The complexity of selecting actions via the \ac{imp} algorithm is the same as the evaluation of the \ac{imp} algorithm on $\mathcal{P}(N, K, m)$ as specified in Section \ref{subsec:complexity}. For $T_{\mathrm{train}}$ training episodes, the total complexity in this part is $\mathcal{O}(T_{\mathrm{train}}M(N-K)(N^2d_{\max} + Nd_{\max}^2))$.

The complexity of reward generation in each episode depends on the decoding algorithm and the achievable \ac{fer} within the training \ac{snr} range $[\gamma_{\min}, \gamma_{\max}]$. Let $\epsilon_{\min}$ represent the achievable \ac{fer} at $\gamma_{\max}$. The number of Monte-Carlo simulations required to generate the reward at each \ac{imp} step is $\mathcal{O}(\epsilon_{\min}^{-1})$, and each Monte-Carlo simulation takes $\mathcal{O}(LN\log N)$ complexity for \ac{cscl} decoding with list size $L$. In total, the complexity of generating rewards is $\mathcal{O}\Paren{T_{\mathrm{train}}\epsilon_{\min}^{-1}(N-K)LN \log N}$.

For the \ac{imp} model optimization, each update takes $\mathcal{O}(Md_{\max}^2)$ computational complexity. Note that this complexity is independent of $N$ because the trainable operations in the \ac{imp} algorithm do not scale with $N$. The complexity of running $T_{\mathrm{train}}$ episodes is $\mathcal{O}(T_{\mathrm{train}}(N-K) Md_{\max}^2)$.

Therefore, the overall complexity of training an \ac{imp} model for $T_{\mathrm{train}}$ episodes on $\mathcal{P}(N,K,m)$ with \ac{cscl} decoding with list size $L$ is $\mathcal{O}\Paren{T_{\mathrm{train}}(N-K)(MN^2 d_{\max} + MNd_{\max}^2 + \epsilon_{\min}^{-1}LN \log N)}$.

When $T_{\mathrm{tune}}$ episodes are used for fine-tuning at design \ac{snr} $\gamma_{\mathrm{eval}}$, an additional training complexity of $\mathcal{O}\Paren{T_{\mathrm{tune}}(N-K)(MN^2 d_{\max} + MNd_{\max}^2 + \epsilon^{-1}LN \log N)}$ is needed at each $\gamma_{\mathrm{eval}}$, where $\epsilon$ is the achievable \ac{fer} at $\gamma_{\mathrm{eval}}$. Note, however, that $T_{\mathrm{tune}}$ is selected as a constant such that $T_{\mathrm{tune}} \ll T_{\mathrm{train}}$.

\section{Experimental Results}\label{sec:simu}

\begin{table}[t]
    \centering
    \begin{tabular}{|>{\centering}m{0.11\textwidth}|>{\centering}m{0.28\textwidth} | m{0.5\textwidth}|}
    \hline
    Algorithm & Evaluation Complexity & Specification and Comments \\ \hline
    Tal-Vardy \cite{tal_construction} & $\cO(N \mu^2 \log \mu)$ & A common choice for the fidelity parameter $\mu$ is $\mu = 2\lfloor \log N \rfloor$, which gives $\cO(N \log^2 N \log \log N)$ complexity. \\ \hline
    5G \cite{3gpp.38.212} & $\cO(1)$ & Only works for $N \le 1024$. \\ \hline
    Nested \cite{Li2021_COML} & $\cO(N)$ & Retraining required for every $(N, \gamma_{\mathrm{eval}})$ combination \\ \hline
    GenAlg \cite{elkelesh_GA} & $\cO(N_{\mathrm{pop}}S \epsilon^{-1} LN\log N)$ & Retraining required for every $(N, K, \gamma_{\mathrm{eval}})$ combination. \\ \hline
    Tabular-RL \cite{liao2021construction} & $\cO(N)$ & Need $\cO(N^3)$ space to store the action-value table. Retraining required for every $(N, K, \gamma_{\mathrm{eval}})$ combination. \\ \hline
    Pure \acs{imp} & $\cO(MN(N-K) (N d_{\max} + d_{\max}^2))$ & A trained \acs{imp} model generalizes to various $(N, K, \gamma_{\mathrm{eval}})$. \\ \hline
    \acs{imp} with NS & $\cO(MN(N-K) (N d_{\max} + d_{\max}^2)$ $ + \epsilon^{-1} L N \log N)$ & A trained \acs{imp} model generalizes to various $(N, K, \gamma_{\mathrm{eval}})$. \\ \hline
    \end{tabular}
    \caption{Evaluation complexity comparison of different polar-code construction methods.}
    \label{tab:eval_complexity}
    \vspace{-2em}
\end{table}

\ac{imp} performance evaluation compares its resultant polar codes' \ac{fer} under \ac{cscl} decoding to polar codes given by 5G NR standard \cite{3gpp.38.212}, Tal-Vardy's algorithm \cite{tal_construction}, the \ac{genalg} \cite{elkelesh_GA}, the nested polar-code construction method \cite{Li2021_COML}, and the tabular \ac{rl} algorithm \cite{liao2021construction} for the \ac{awgn} channel with \ac{bpsk} modulation. Table~\ref{tab:eval_complexity} lists the complexity of constructing a $\mathcal{P}(N, K, m)$ polar code by each of these algorithms. For each evaluation point, the number of simulated transmissions is such that the number of observed frame errors is at least $500$ and the number of total simulated transmissions is at least $10^6$. The hyper-parameters adopted in the implementation of Tal-Vardy, \ac{genalg}, the tabular \ac{rl} method, and the \ac{imp} algorithm are:

\begin{enumerate}
    \item \textit{Tal-Vardy} \cite{tal_construction}: The fidelity parameter is set as $\mu = 2 \lfloor \log N\rfloor$.
    \item \textit{\ac{genalg}} \cite{elkelesh_GA}: The hyper-parameters are consistent with the ones adopted in \cite{elkelesh_GA}: the population size is $S=20$, and the number of truncated parents is $T=5$. The number of Monte-Carlo simulations for each iteration's \ac{fer} estimates is set such that either the number of observed frame errors reaches $10^3$ or the number of simulated frames reaches $10^6$. The maximum number of population update iterations is $N_{pop} = 1000$. 
    \item \textit{Tabular \ac{rl} method} \cite{liao2021construction}: The discount rate is set to $1$. The trace decay factor $\lambda$, the neighborhood update rate $\kappa$, and the exploration rate $\epsilon$ for the $\epsilon$-greedy policy are initialized as $\lambda = 0.8$, $\kappa = 1$, and $\epsilon = 0.5$, respectively. As the training proceeds, $\lambda$, $\kappa$, and $\epsilon$ are updated gradually towards $0.9$, $0$, and $\frac{1}{5N}$ for length-$N$ polar codes, respectively. The maximum number of training episodes is $2 \times 10^5$.
    \item \textit{\ac{imp} algorithm}: The maximum message passing iterations $M$ is set to $3$. For initialization operations, $d_{\mathrm{loc}} = 4$ and $d_{\mathrm{type}} = 28$, so $d^{(0)} = 32$. After each message passing iteration, is $d^{(i)} = 64$, $i \in\{1,2,3\}$. The global features $\bm{g}_C$ and $\bm{g}_V$ both have dimension $d_{\mathrm{pool}}=1$. The post-processing \ac{mlp} has two hidden layers with size $[128, 32]$, respectively. \changed{With these specifications, the \ac{imp} model contains $7.5 \times 10^4$ real-valued trainable coefficients.} The \ac{crc} generator polynomial remains constant during training and evaluation. For the reward generation, $P_{e}\Paren{\mathcal{I}_{N,t}, N, m, L, \gamma}$ and  $P_{e}\Paren{\mathcal{I}_{N,{t+1}}, N, m, L, \gamma}$ in \eqref{eq:instant_reward} are estimated at each step $t$ by a Monte-Carlo simulation with at most $100$ observed frame errors and at most $10^5$ frames in total. The \ac{dql} algorithm with a replay buffer of size $10^4$ and a target Q-value network \cite{mnih2013playing} that updates every $2$ episodes is used. During training, the exploration rate is initialized as $0.5$, and exponentially decayed to $\frac{1}{5N}$ with $0.999$ decay rate. The discount factor $\beta$ is initialized as $0.8$, which increases linearly to $1$ in $20$ episodes and remains at $1$ afterwards. The \ac{imp} models are trained for $T_{\mathrm{train}}=10^5$ episodes on $N=64$ cases, and $T_{\mathrm{train}}=5\times 10^5$ episodes on $N=128$ cases, respectively, before fine-tuning. The number of additional episodes for fine-tuning the \ac{imp} model is $T_{\mathrm{tune}}=100$.
\end{enumerate}

The remainder of this section evaluates the \ac{imp} algorithm in three aspects: Section~\ref{subsec:simu_fer} shows that \ac{imp}-based polar codes outperform the state-of-the-art constructions in many cases when the \ac{imp} model is fine-tuned for the evaluation case; Section~\ref{subsec:simu_gen_snr} verifies the generalization capability of a trained \ac{imp} model to different design \acp{snr} and list sizes; and Section~\ref{subsec:simu_gen_N} illustrates that a trained \ac{imp} model directly applies to polar-code construction tasks with different rates and blocklengths, and yet still provides good polar codes.

\subsection{Error-Correction Performance}\label{subsec:simu_fer}

\begin{figure}[t!]
    \centering
    \includegraphics[width=0.88\textwidth]{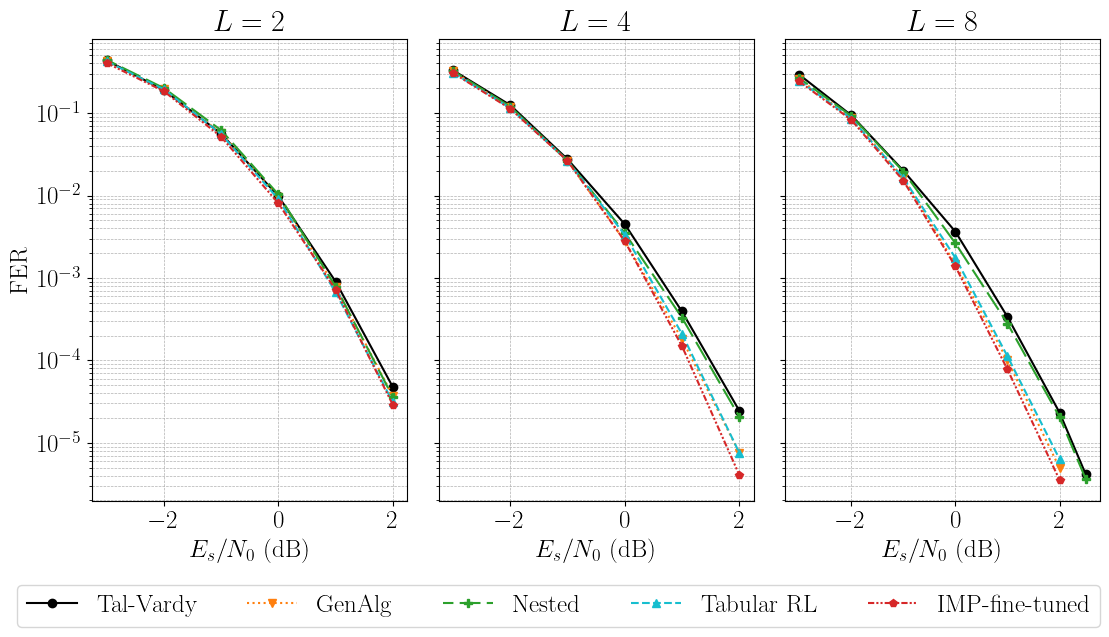}
    \caption{\acp{fer} of $\mathcal{P}(64, 32, 4)$ given by different polar-code construction methods under \ac{cscl} decoding with list sizes $L=2$ (left); $4$ (middle); $8$ (right). The \ac{crc} polynomial is 0x3.}
    \label{fig:64_32_4_opt}
    \vspace{-1em}
\end{figure}

\begin{figure}[t!]
    \centering
    \includegraphics[width=0.88\textwidth]{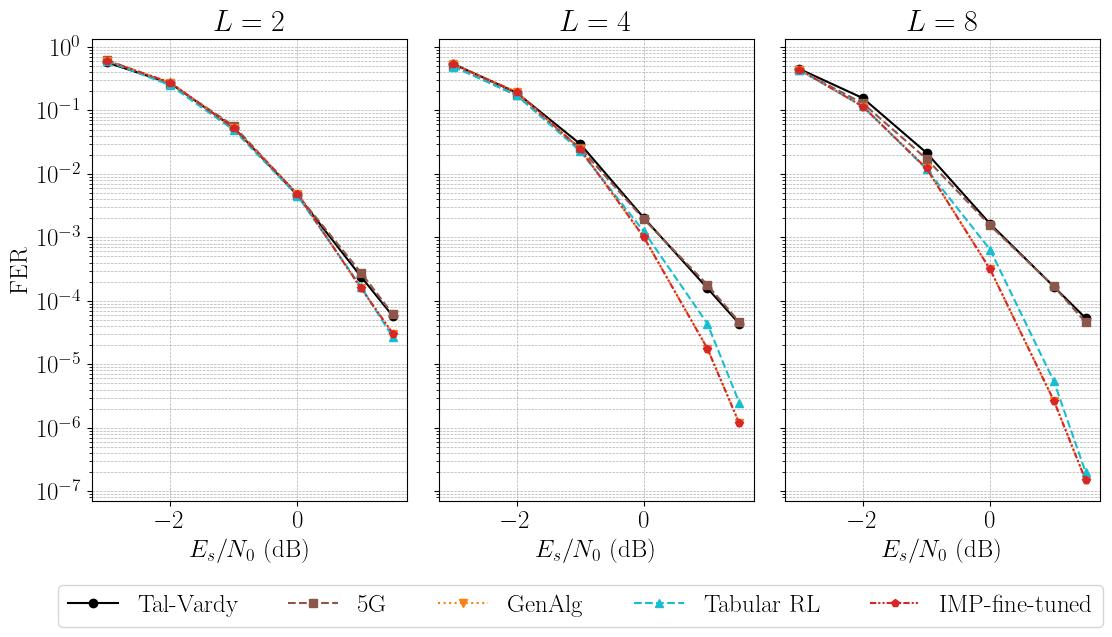}
    \caption{\acp{fer} of $\mathcal{P}(128, 64, 4)$ given by different polar-code construction methods under \ac{cscl} decoding with list sizes $L=2$ (left); $4$ (middle); $8$ (right). The \ac{crc} polynomial is 0x3.}
    \label{fig:128_64_4_opt}
    \vspace{-1em}
\end{figure}

Figs.~\ref{fig:64_32_4_opt} and \ref{fig:128_64_4_opt} compare the \ac{fer} of the learned constructions by the \ac{imp} algorithm with pointwise fine-tuning to the constructions given by the Tal-Vardy's method \cite{tal_construction}, the 5G NR standard \cite{3gpp.38.212}, the \ac{genalg} \cite{elkelesh_GA}, the nested polar-code construction method \cite{Li2021_COML}, and the tabular \ac{rl} method \cite{liao2021construction} for $\mathcal{P}(64, 32, 4)$ and $\mathcal{P}(128, 64, 4)$, respectively. At each $\gamma_{\mathrm{eval}}$, a polar code tailored for design \ac{snr} $\gamma_{\mathrm{eval}}$ is generated by each of the construction methods\footnote{\cite{Li2021_COML} only provided the nested polar code for $N=64$ at $\text{FER}=10^{-2}$. The non-frozen set selection for $N=64$, $K=32$ case in \cite{Li2021_COML} is adopted here for evaluation.}, and then evaluated by \ac{cscl} decoding at the same \ac{snr} $\gamma_{\mathrm{eval}}$. Since the constructions are channel-dependent, the polar codes corresponding to different $\gamma_{\mathrm{eval}}$ are potentially different, even when they are generated by the same construction method. The training \ac{snr} before fine-tuning is selected uniformly at random from the range $[-3, 0.5]$ dB in each episode for $\cP(64, 32, 4)$, and the corresponding training \ac{snr} range is $[-2.5, 0.5]$ dB for $\cP(128, 64, 4)$.

\begin{figure}[t!]
    \centering
    \includegraphics[width=0.5\textwidth]{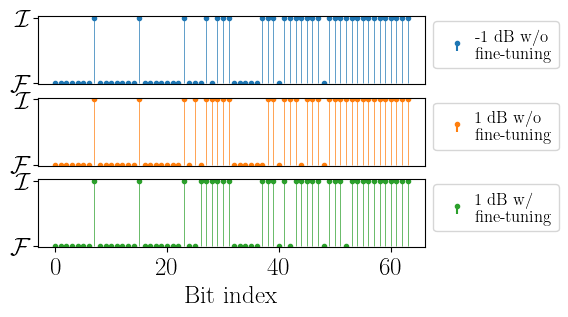}
    \caption{$\cP(64, 32, 4)$ constructed by the ``\ac{imp}-$[-3.0, 0.5]$-L$4$'' model evaluated at $\gamma_{\mathrm{eval}} = -1$ dB (top) and $\gamma_{\mathrm{eval}} = 1$ dB (middle), and $\cP(64, 32, 4)$ constructed by the ``\ac{imp}-fine-tuned'' model at $\gamma_{\mathrm{eval}} = 1$ dB (bottom).}
    \label{fig:64_32_constructions}
    \vspace{-2em}
\end{figure}

Figs.~\ref{fig:64_32_4_opt} and \ref{fig:128_64_4_opt} show that the constructions from the \ac{imp}-fine-tuned method outperform Tal-Vardy's and 5G constructions in all evaluated scenarios under \ac{cscl} decoding with various list sizes $L$. The improvement in \ac{fer} increases with $L$. Figs.~\ref{fig:64_32_4_opt} and \ref{fig:128_64_4_opt} also show that the polar codes constructed by the \ac{imp}-fine-tuned model achieve the lowest \ac{fer} among codes generated by these considered benchmark methods tailored for \ac{cscl} decoding. For a given $\gamma_{\mathrm{eval}}$ and a given decoding list size $L$, the \ac{imp} algorithm with fine-tuning finds the same $\cP(128, 64, 4)$ as the \ac{genalg} does. 
When the two algorithms identify the same polar codes at a given $\gamma_{\mathrm{eval}}$, the \ac{imp} algorithm still shows its advantage over the \ac{genalg} in the sense that the $\gamma_{\mathrm{eval}}$ values that are not seen during training can be directly fed into the same trained \ac{imp} model to generate corresponding polar codes. In contrast, re-training is needed for the \ac{genalg} model when $\gamma_{\mathrm{eval}}$ deviates from the design \ac{snr}.

\subsection{Generalization in Design \acp{snr} and List Sizes} \label{subsec:simu_gen_snr}

This section illustrates the \ac{imp} model's generalization to various design \acp{snr} and to different list sizes without fine-tuning. Since the target \ac{snr} is one of \ac{imp} model's input features, a single trained \ac{imp} model can automatically provide different constructions at different target \acp{snr}. For example, Fig.~\ref{fig:64_32_constructions} shows that the two polar codes generated by the same ``\ac{imp}-$[-3.0, 0.5]$-L$4$'' model evaluated at $\gamma_{\mathrm{eval}} = -1$ dB and $\gamma_{\mathrm{eval}} = 1$ dB are distinct.

\begin{figure}[t!]
    \centering
    \begin{subfigure}[t]{0.46\textwidth}
        \centering
        \includegraphics[width=\textwidth]{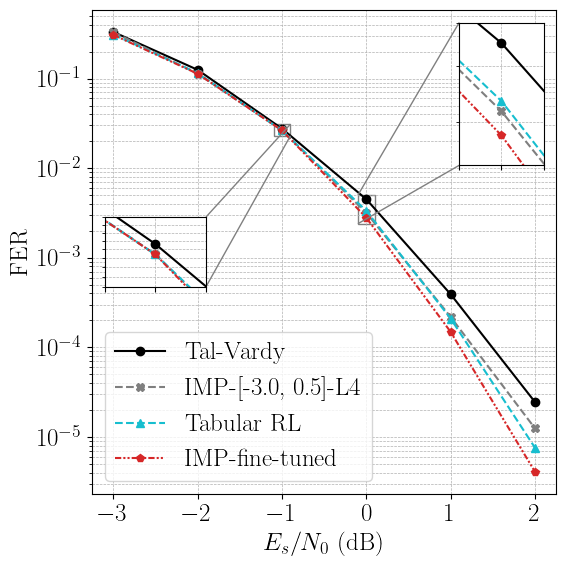}
        \caption{\ac{fer} of $\cP(64, 32, 4)$ under \ac{cscl} decoding with $L=4$. The degree-4 \ac{crc} polynomial is 0x3.}
        \label{subfig:64_32_4_range_L4}
    \end{subfigure}\hspace{1em}
    \begin{subfigure}[t]{0.46\textwidth}
        \centering
        \includegraphics[width=\textwidth]{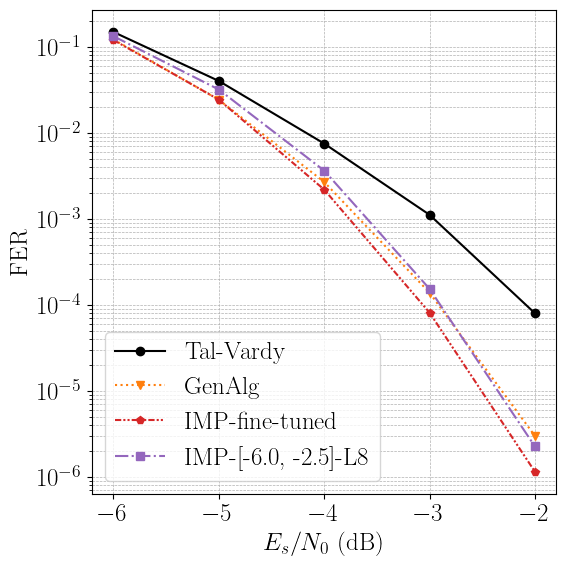}
        \caption{\ac{fer} of $\cP(128, 32, 4)$ under \ac{cscl} decoding with $L=8$. The degree-4 \ac{crc} polynomial is 0x3.}
        \label{subfig:128_32_4_range}
    \end{subfigure}
    \caption{Comparison of polar codes constructed by the \ac{imp} model with and without fine-tuning.}
    \label{fig:64_32_4_range}
    \vspace{-2em}
\end{figure}

Fig.~\ref{subfig:64_32_4_range_L4} shows constructions of $\cP(64, 32, 4)$ for \ac{cscl} decoding with $L=4$, in which the ``\ac{imp}-fine-tuned'' models use the ``\ac{imp}-$[-3.0, 0.5]$-L$4$'' model as the starting point of fine tuning. Fig.~\ref{subfig:64_32_4_range_L4} shows that the \ac{imp} model trained over the \ac{snr} range $[-3, 0.5]$ dB without fine-tuning can generate constructions that outperform the Tal-Vardy's constructions over a large range of design \acp{snr}, even when $\gamma_{\mathrm{eval}}$ is outside the training \ac{snr} range of $[-3, 0.5]$ dB. Also, at some evaluation points, e.g., $\gamma_{\mathrm{eval}} = -1$ dB, the \ac{imp} model without fine-tuning already finds the same construction as the tabular \ac{rl} and as the fine-tuned model, while the latter two are trained specifically for that single $\gamma_{\mathrm{eval}}$ point. These observations indicate that the \ac{imp} model learns the general rules for constructing polar codes tailored for a given \ac{cscl} decoder, and that these learned rules can be applied to a wide range of design \acp{snr}, even to \acp{snr} outside the training \ac{snr} range. On the other hand, Fig.~\ref{subfig:64_32_4_range_L4} shows that the additional fine-tuning at each evaluation point can effectively improve the learned construction in most cases, especially when the evaluation point is outside the \ac{imp} model's initial training range as evident in the bottom two cases in Fig.~\ref{fig:64_32_constructions}.

Similar observations follow from Fig.~\ref{subfig:128_32_4_range} for $\cP(128, 32, 4)$ under \ac{cscl} decoding with $L=8$. Specifically, the \ac{imp} model without fine-tuning outperforms Tal-Vardy's method, and fine-tuning can further improve the constructed codes. However, in Fig.~\ref{subfig:128_32_4_range}, a strictly positive gap is observed between the performance of the trained \ac{imp} model over the entire \ac{snr} range and the pointwise fine-tuned models, indicating that the trained \ac{imp} model is strictly sub-optimal without fine-tuning even within the training \ac{snr} range. This gap is mainly caused by the wide training \ac{snr} range of $[-6, -2.5]$ dB, in which the \ac{fer} spans about five orders of magnitude. Due to the wide \ac{fer} span, the current \ac{imp} model fails to generate accurate priority metrics for all training scenarios. This suggests that a more expressive \ac{nn} architecture might improve the \ac{imp} model when fitting a single \ac{imp} model to a wide \ac{snr} (and consequently \ac{fer}) range (e.g., several orders of \ac{fer} magnitude).

\begin{figure}[t!]
    \centering
    \includegraphics[width=0.5\textwidth]{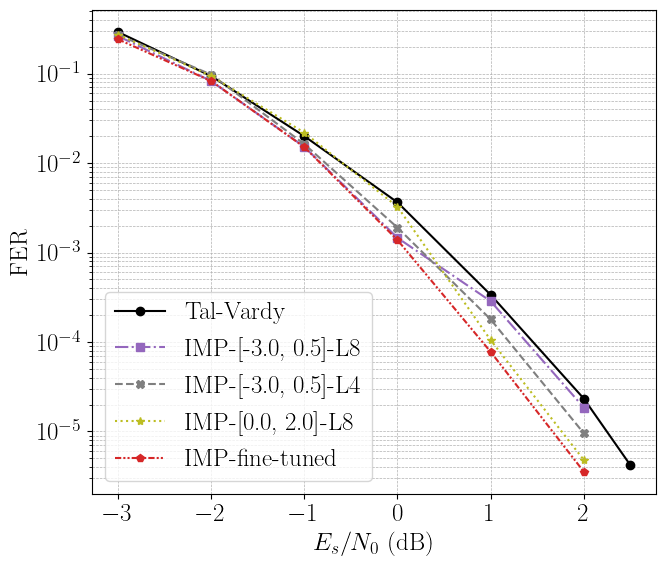}
    \caption{\ac{fer} performance of $\mathcal{P}(64, 32, 4)$ given by different polar-code construction methods under \ac{cscl} decoding with $L=8$. The degree-4 \ac{crc} polynomial is 0x3.}
    \label{fig:64_32_4_range_L8}
    \vspace{-2em}
\end{figure}

Fig.~\ref{fig:64_32_4_range_L8} compares different construction methods for $\cP(64, 32, 4)$ under \ac{cscl} decoding with $L=8$. The ``\ac{imp}-$[-3.0, 0.5]$-L$8$'' model is fine tuned to obtain the ``\ac{imp}-fine-tuned'' models. Both ``\ac{imp}-$[-3.0, 0.5]$-L$8$'' and ``\ac{imp}-$[0.0, 2.0]$-L$8$'' models provide constructions that outperform Tal-Vardy's constructions over the entire evaluation \ac{snr} range, though the improvement is minimal for both models when $\gamma_{\mathrm{eval}}$ is outside the training ranges. Again, fine-tuning helps to find good codes with even lower \acp{fer}. Fig.~\ref{fig:64_32_4_range_L8}'s \ac{fer} curves for the model ``\ac{imp}-$[-3.0, 0.5]$-L$8$'' and the model ``\ac{imp}-$[0.0, 2.0]$-L$8$'' intersect between $0$ dB and $1$ dB \ac{snr}. This implies that if the \ac{imp} model is applied without fine-tuning, better constructions are expected when the design \ac{snr} is within the training \ac{snr} range. 

Fig.~\ref{fig:64_32_4_range_L8} also shows the effect of decoding scheme mismatch during training and evaluation. Specifically, under the \ac{cscl} decoding with $L=8$, the \ac{fer} of polar codes constructed by the ``\ac{imp}-$[-3.0, 0.5]$-L$4$'' model is contrasted with that of polar codes constructed by the ``\ac{imp}-$[-3.0, 0.5]$-L$8$'' model. With the same training \ac{snr} range, the polar codes constructed by the latter \ac{imp} model have lower \ac{fer} compared to the polar codes constructed by the former one when evaluated within the training \ac{snr} range. This suggests that the constructions given by the trained \ac{imp} model is effectively tailored for the decoding scheme during training. On the other hand, the codes constructed by the ``\ac{imp}-$[-3.0, 0.5]$-L$4$'' model show constantly lower \acp{fer} than the Tal-Vardy's constructions under the \ac{cscl} decoding with $L=8$, meaning that the \ac{imp} models trained for one \ac{cscl} decoding scheme likely provide constructions that are also reasonably good in \ac{fer} performance under \ac{cscl} decoding with different list sizes.

\subsection{Scalability in $N$ and $K$}\label{subsec:simu_gen_N}

\begin{figure}
    \centering
    \includegraphics[width=0.8\textwidth]{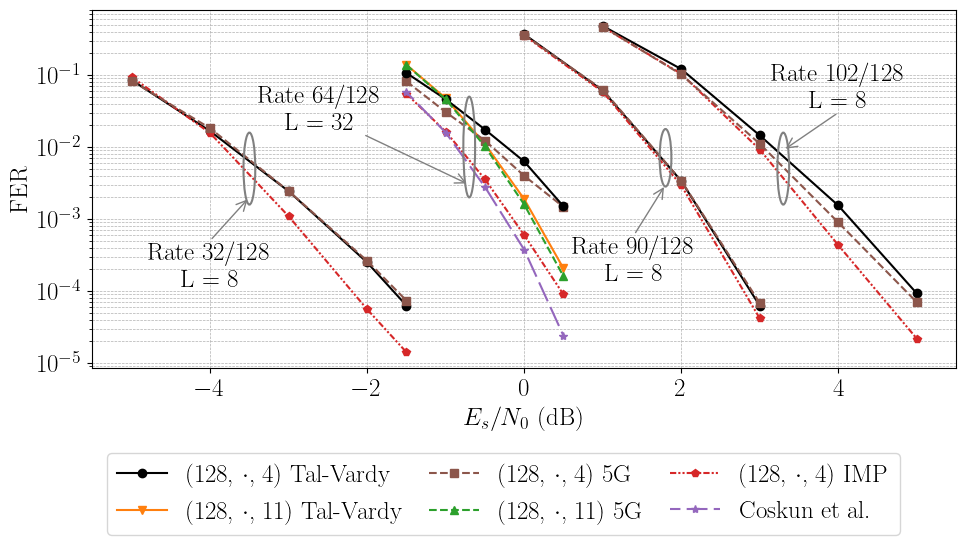}
    \caption{Generalization of ``\ac{imp}-$[-3.0, 0.5]$-L$4$-N$128$-K$64$'' model to $N=128$ and different values of $K$. The Tal-Vardy, 5G, and \ac{imp} constructions are evaluated by \ac{cscl} decoding. \ac{scl} decoding is used to evaluate the performance of dRM \cite{cocskun2020successive} and the polar codes given by Co{\c s}kun \emph{et al.} in \cite{Coskun2022}. A 4-bit \ac{crc} with polynomial 0x3 is used in all four cases for the \ac{imp} constructions, and in the $L=8$ cases for the Tal-Vardy and 5G constructions. For the rate-$1/2$, $L=32$ case, 5G \ac{crc}-11 is adopted to evaluate Tal-Vardy and 5G constructions. }
    \label{fig:128_high_rate}
    \vspace{-2em}
\end{figure}

Fig.~\ref{fig:128_high_rate} applies directly the ``\ac{imp}-$[-3.0, 0.5]$-L$4$'' model trained on $\cP(128, 64, 4)$ to construct polar codes with $N=128$ and various values of $K$ without additional training. The \ac{imp} model is relabeled as ``\ac{imp}-$[-3.0, 0.5]$-L$4$-N$128$-K$64$'' for better clarity.

For the rate-$1/2$ case, the constructions of $\cP(128, 68, 4)$ given by the ``\ac{imp}-$[-3.0, 0.5]$-L$4$-N$128$-K$64$'' model is compared against $\cP(128, 68, 4)$ and $\cP(128, 75, 11)$ given by the 5G ordering and Tal-Vardy's method under \ac{cscl} decoding with $L=32$. In particular, for Tal-Vardy's method and the \ac{imp} algorithm, the polar codes are constructed separately at each $\gamma_{\mathrm{eval}}$. \changed{The performance of the polar code with $N=128$ and $K=64$ given by Co{\c s}kun \emph{et al.} in \cite{Coskun2022} under \ac{scl} with $L=32$ is also included for comparison, in which dynamic frozen bits are utilized.} As Fig.~\ref{fig:128_high_rate} depicts, the \ac{imp}-based constructions clearly outperform the constructions given by 5G and Tal-Vardy, even when a longer \ac{crc} is used to aid the decoding for the latter two methods. In comparison to the polar-code designs with dynamic frozen bits, the \ac{imp}-based constructions decoded by \ac{cscl} with $4$-bit \ac{crc} show similar performance in the low \ac{snr} regime, while the dynamic-frozen-bit-enabled polar codes achieve lower \ac{fer} than the \ac{imp}-based codes in the high \ac{snr} regime.

Besides the rate-$1/2$ case, the constructions of $\cP(128, 36, 4)$, $\cP(128, 94, 4)$ and $\cP(128, 106, 4)$ given by 5G, Tal-Vardy, and the ``\ac{imp}-$[-3.0, 0.5]$-L$4$-N$128$-K$64$'' model are compared under \ac{cscl} decoding with $L=8$. The \ac{imp}-based constructions outperform the corresponding polar codes specified by 5G and Tal-Vardy's algorithm for all these three $K$ values.

Note that no additional training is included in the \ac{imp} construction of polar codes with different values of $K \neq 64$. The observation that such \ac{imp}-based constructions outperform the 5G and Tal-Vardy's construction schemes, and that these constructions achieve comparable performance to the polar code designs with dynamic frozen bits indicate \ac{imp}'s good generalization capability to different $K$ values. In other words, a single trained \ac{imp} model directly finds polar codes with various code rates that achieve reasonably good performance under \ac{cscl} decoding.

\begin{figure}[t!]
    \centering
    \begin{subfigure}[t]{0.49\textwidth}
    \vskip 0pt
            \includegraphics[width=\textwidth]{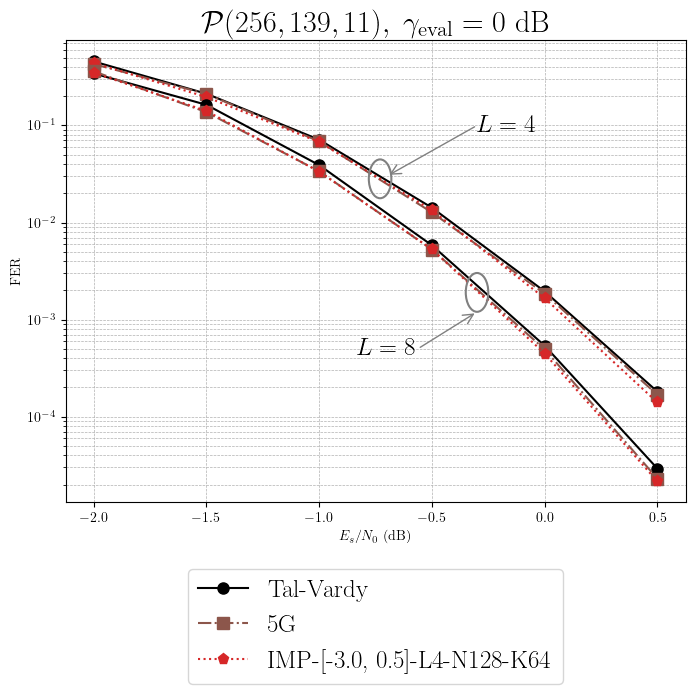}
            \caption{$\mathcal{P}(256, 139, 11)$ codes with design \ac{snr} at $0$ dB.}
            \label{fig:256_139}
    \end{subfigure}%
    \begin{subfigure}[t]{0.49\textwidth}
    \vskip 0pt
        \includegraphics[width=\textwidth]{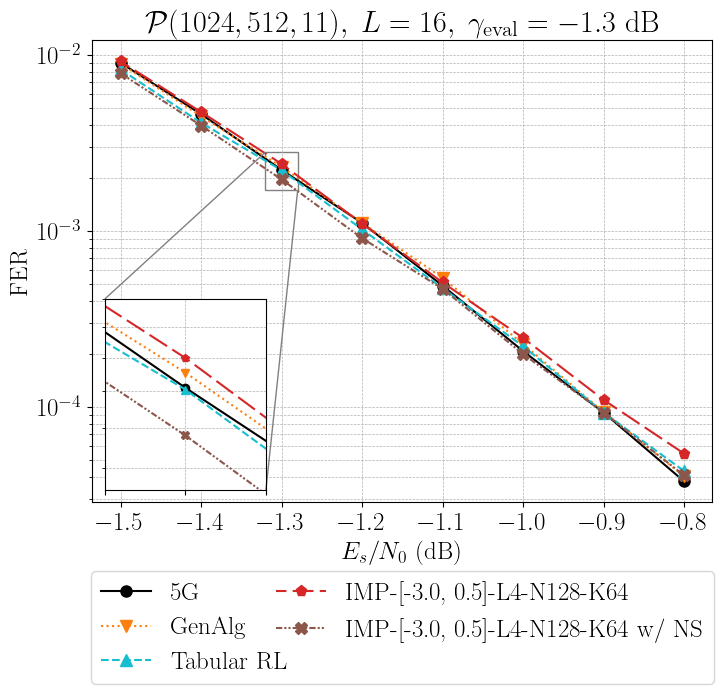}
        \caption{$\mathcal{P}(1024, 512, 11)$ codes with design \ac{snr} at $-1.3$ dB.}
    \label{fig:1024_512_11_1.7}
    \end{subfigure}
    \caption{\ac{fer} performance of applying a trained \ac{imp} model to different $N$'s without fine-tuning. The 5G \ac{crc}-11 with polynomial 0x621 is adopted in evaluation.}
    \label{fig:different_N}
    \vspace{-2em}
\end{figure}

\changed{Fig.~\ref{fig:different_N} illustrates \ac{imp}'s scalability by using the trained ``\ac{imp}-$[-3.0, 0.5]$-L$4$-N$128$-K$64$'' model to construct polar codes with larger blocklengths $N=256$ and $1024$. The design \ac{snr} for all methods is fixed at $\gamma_{\mathrm{eval}} = 0$ dB for $N=256$ case, and $\gamma_{\mathrm{eval}} = -1.3$ dB for $N=1024$ case, respectively, and the constructed codes are evaluated over the entire \ac{snr} range.} 
The polar-code construction for the curve labeled ``\ac{imp}-$[-3.0, 0.5]$-L$4$-N$128$-K$64$'' is obtained by feeding \changed{the corresponding $N$, $K$, $\gamma_{\mathrm{eval}}$ values} to the \ac{imp} model, which is trained only on $N=128$, $K=64$. 
\changed{For the $N=256$ case in Fig.~\ref{fig:256_139}, applying the trained \ac{imp} model immediately provides polar codes with slightly better performance compared to Tal-Vardy's and 5G polar code under \ac{cscl} decoding with $L=4$ and $L=8$, both at $\gamma_{\mathrm{eval}}$ and at various other evaluated \ac{snr} values.}
\changed{For the $N=1024$ case as shown in Fig.~\ref{fig:1024_512_11_1.7}, the direct generalization of the trained \ac{imp} model shows similar performance comparing to the 5G construction. The \ac{snr} gap at \ac{fer} $2 \times 10^{-3}$ between the \ac{imp} construction and both the 5G polar code and the tabular \acs{rl} construction is $0.01$ dB; the corresponding gap between the \ac{imp} and \ac{genalg} constructions is less than $0.005$ dB, while both tabular \ac{rl} and \ac{genalg} methods are trained on $N=1024$, $K=512$, $\gamma=-1.3$ dB directly.}

Further improvement \changed{in the $N=1024$ case} occurs by adding an additional \ac{ns} over the input \acp{snr}, i.e., the values of $x_u$ for all variable nodes $u \in \mathcal{Y}_N$, to the \ac{imp} model. In particular, five \ac{snr} values $\{-1.5, -1.4, -1.3, -1.2, -1.1\}$ are fed into the trained \ac{imp} model to generate five candidate constructions. These candidates are then evaluated with \ac{cscl} decoding and $\text{SNR} = -1.3$ dB, and the construction that achieves the lowest \ac{fer} is selected as the final outcome. Note that the \ac{ns} process evaluates a fixed trained \ac{imp} model for several times, and does not change the trainable parameters of the model. In other words, no training is included in \ac{ns}. The complexity of \ac{ns} for $t_{\mathrm{ns}}$ neighboring \ac{snr} values is $\mathcal{O}(t_{\mathrm{ns}}M(N-K)(N^2 d_{\max} + Nd_{\max}^2) + t_{\mathrm{ns}}\epsilon^{-1}LN \log N)$, where $\epsilon$ is the achievable \ac{fer} at the design \ac{snr}.

Fig.~\ref{fig:1024_512_11_1.7} shows that \ac{ns} helps find constructions with lower \ac{fer} at $-1.3$ dB \ac{snr}. More specifically, the polar-code construction found with \ac{ns} outperforms 5G polar code, the tabular \ac{rl} construction, and the \ac{genalg} construction at the design \ac{snr} $-1.3$ dB. The adopted \ac{imp} model, in particular, applies with no additional training on $N$, $K$, and $\gamma_{\mathrm{eval}}$. These observations verify the \ac{imp} model's scalability to different blocklengths without additional training.

\begin{figure}[t!]
    \centering
    \includegraphics[width=0.9\textwidth]{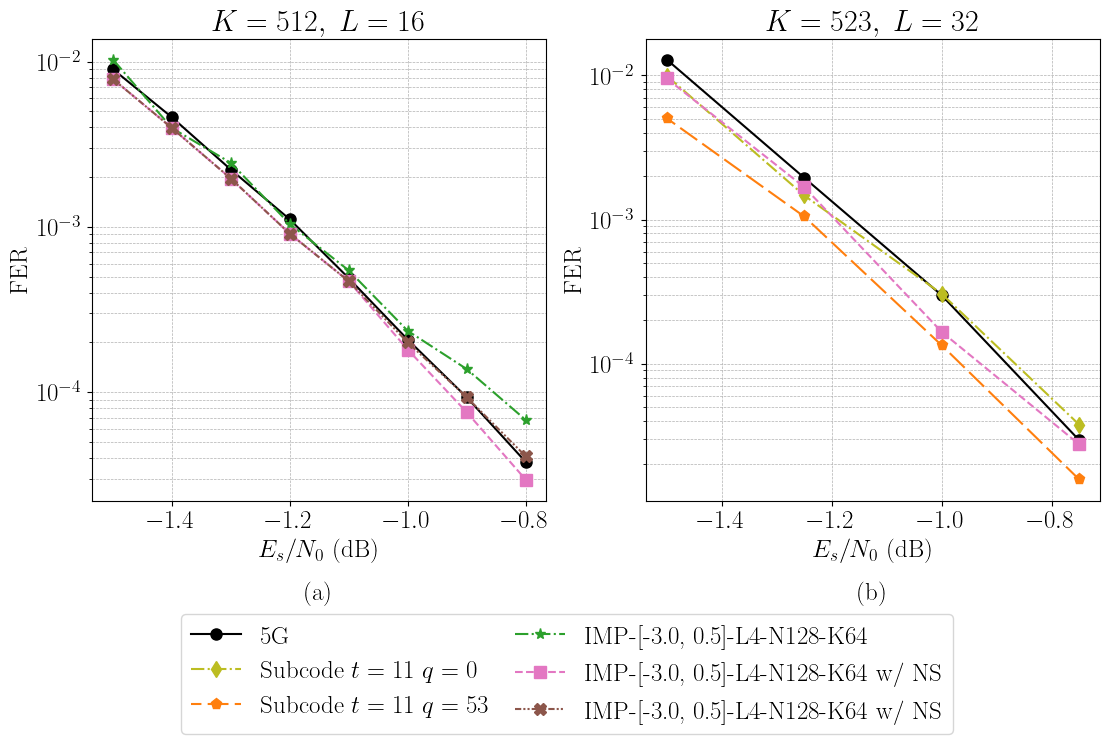}
    \caption{(a) Effect of \ac{ns} on $\mathcal{P}(1024, 512, 11)$ under \ac{cscl} decoding with $L=16$. (b) Performance of \ac{imp} with \ac{ns} on $\mathcal{P}(1024, 523, 11)$ under \ac{cscl} decoding with $L=32$. }
    \label{fig:1024_512_11_NS}
    \vspace{-2em}
\end{figure}

Fig.~\ref{fig:1024_512_11_NS} further shows \ac{ns}'s effect on the \ac{imp} model for blocklength $N=1024$ that is unseen during training of the ``\ac{imp}-$[-3.0, 0.5]$-L$4$-N$128$-K$64$'' model. $\mathcal{P}(1024, 512, 11)$ and $\mathcal{P}(1024, 523, 11)$ are evaluated. The curve labeled ``\ac{imp}-$[-3.0, 0.5]$-L$4$-N$128$-K$64$'' shows the \ac{imp} polar codes' performance when the model's input \ac{snr} matches the evaluation \ac{snr}, while the curve labeled ``\ac{imp}-$[-3.0, 0.5]$-L$4$-N$128$-K$64$ w/ \ac{ns}'' reports the performance of the polar-code constructions after \ac{ns} at each evaluation \ac{snr}.

In Fig.~\ref{fig:1024_512_11_NS}a, the direct generalization of a trained \ac{imp} model without \ac{ns} can generate polar codes that achieve the same order of \ac{fer} magnitude as the 5G polar codes. This indicates that the \ac{imp} model learns some general polar-code-construction rules that apply to various blocklengths. Nevertheless, the \ac{fer} of \ac{imp} polar codes without \ac{ns} may be higher than that of the 5G polar code, and there is no guarantee or prior knowledge of whether the \ac{imp} model without \ac{ns} provides a satisfying polar-code construction for a target $(N,K,\gamma)$ when the target blocklength $N$ is never seen by the \ac{imp} model during training. The lack of performance guarantee in generalization is a major limitation of the current \ac{imp} algorithm. \ac{ns} is used as a simple remedy for this limitation that requires no additional training. As Fig.~\ref{fig:1024_512_11_NS}a shows, in many cases, a polar-code construction with a lower \ac{fer} can be found by evaluating several neighboring \ac{snr} points, and the reported polar codes after \ac{ns} can outperform the 5G polar code.

In Fig.~\ref{fig:1024_512_11_NS}b, the constructions given by \ac{imp} with \ac{ns} for the $K=523$ case are compared against the randomized polar subcode~\cite{trifonov2017randomized}, which exploits constructions with dynamic frozen bits \changed{and uses the sequential decoder}. The parameter $t$ in \cite{trifonov2017randomized} represents the number of dynamic frozen bits that are dependent on previous information bits, and $q$ represents the number of dynamic (random) freezing constraints, \changed{so the design exploits $t+q$ dynamic frozen bits in total}. To achieve a fair comparison, this work selects $t=11$ to match the \ac{crc} length, \changed{and the sequential decoder's list size $L=32$ to match the \ac{cscl} decoder's list size.} The figure again shows that \ac{imp} with \ac{ns} finds constructions that achieve lower \ac{fer} than 5G polar code, and these \ac{imp}-with-\ac{ns} constructions show better performance comparing to the randomized polar subcode when $t$ matches the \ac{crc} length and $q=0$. The randomized polar subcode outperforms \ac{imp} with \ac{ns} when $q=53$. This is expected because the randomized polar subcode allows more flexibility in the frozen-bit values and needs a more sophisticated \changed{list} decoder.

\vspace{-1em}
\section{Conclusion}
\label{sec:conc}
This paper proposes a \ac{gnn}-based polar-code construction algorithm, named the \ac{imp} algorithm. A salient feature of the \ac{imp} algorithm is that a single trained \ac{imp} model directly applies to constructions for various design \acp{snr} and different blocklengths without any additional training. This feature makes the \ac{imp} algorithm a powerful candidate in real-world deployment, in which the wireless channel condition varies over time and searching for good polar codes for each design requirement separately can be complicated and costly. 

There exist some limitations in the current \ac{imp} algorithm such as the lack of a general performance guarantee to scenarios that are not seen by the model during training and the high evaluation complexity, and these merit further investigation. Nonetheless, the \ac{imp} algorithm illustrates the potential of using \ac{gnn} as a tool for code design problems. Some future research directions using similar methods may include: (a) polar-code constructions on different channels such as fading channels; (b) variations of the graph structures and initial local messages, e.g., different connection patterns among check nodes, and using Bhattacharyya parameters as the check nodes' initial messages, etc; (c) graph-based algorithms that enable joint learning over non-frozen set and \ac{crc} design; and (d) \ac{gnn}-based code designs tailored for other decoding algorithms such as \ac{bp} decoding, or for other outer codes.

\balance
\bibliographystyle{IEEEtran}
\bibliography{IEEEabrv,references}
\end{document}